\documentclass[aps,nofootinbib,preprintnumbers,twocolumn,superscriptaddress,prd]{revtex4}

\usepackage{amsmath}
\usepackage{amssymb}
\usepackage{amsfonts}
\usepackage{dsfont}
\usepackage{slashed}
\usepackage{graphicx}
\usepackage{graphics}
\usepackage{epsfig}
\usepackage{bbm,bm}
\usepackage{psfrag}
\usepackage{hyperref}
\usepackage{color}
\usepackage{ulem}

\graphicspath{{figures/}}

\definecolor{comment}{rgb}{0.9,0,0}

\newcommand{\pslash}{p \hspace{-0.15cm}/}

\newcommand{\be}{\begin{eqnarray}}
\newcommand{\ee}{\end{eqnarray}}

\newcommand{\Nf}{N_{\text{f}}}
\newcommand{\NL}{N_{\text{L}}}
\newcommand{\Ng}{N_{\text{g}}}

\newcommand{\pat}{\partial_t}
\newcommand{\Eqref}[1]{Eq.~\eqref{#1}}
\newcommand{\dG}{d_{\text{ad}}}

\newcommand{\spFP}{\mathcal{A}}

\begin{document}

\title{An asymptotic safety scenario for gauged chiral Higgs-Yukawa models}

\author{Holger Gies}
\email{holger.gies@uni-jena.de}
\affiliation{\mbox{\it Theoretisch-Physikalisches Institut, Friedrich-Schiller-Universit{\"a}t Jena,}
\mbox{\it D-07743 Jena, Germany}}

\author{Stefan Rechenberger}
\email{rechenbe@uni-mainz.de}
\affiliation{\mbox{\it Institut f{\"u}r Physik, Johannes-Gutenberg-Universit{\"a}t Mainz,}
\mbox{\it D-55099 Mainz, Germany}}
\affiliation{\mbox{\it Faculty of Science (IMAPP), Radboud University Nijmegen,}
\mbox{\it NL-6500 GL Nijmegen, The Netherlands}}

\author{Michael M. Scherer}
\email{m.scherer@thphys.uni-heidelberg.de}
\affiliation{\mbox{\it 	Institut f\"ur Theoretische Physik,
Universit\"at Heidelberg,}
\mbox{\it D-69120 Heidelberg, Germany}}

\author{Luca Zambelli}
\email{luca.zambelli@uni-jena.de}
\affiliation{\mbox{\it Theoretisch-Physikalisches Institut, Friedrich-Schiller-Universit{\"a}t Jena,}
\mbox{\it D-07743 Jena, Germany}}
\affiliation{\mbox{\it Dip. di Fisica, Universit\`a degli Studi di Bologna,}
\mbox{\it INFN Sez. di Bologna
via Irnerio 46, I-40126 Bologna, Italy}}



\begin{abstract} 

  We investigate chiral Higgs-Yukawa models with a non-abelian gauged
  left-handed sector reminiscent to a sub-sector of the standard
  model. We discover a new weak-coupling fixed-point behavior that
  allows for ultraviolet complete RG trajectories which can be
  connected with a conventional long-range infrared behavior in the
  Higgs phase. This non-trivial ultraviolet behavior is characterized
  by asymptotic freedom in all interaction couplings, but a quasi
  conformal behavior in all mass-like parameters. The stable
  microscopic scalar potential asymptotically approaches flatness in
  the ultraviolet, however, with a non-vanishing minimum increasing
  inversely proportional to the asymptotically free gauge
  coupling. This gives rise to nonperturbative -- though weak-coupling
  -- threshold effects which induce ultraviolet stability along a line
  of fixed points.  Despite the weak-coupling properties, the system
  exhibits non-Gau\ss ian features which are distinctly different from
  its standard perturbative counterpart: e.g., on a branch of the line
  of fixed points, we find linear instead of quadratically running
  renormalization constants. Whereas the Fermi constant and the top
  mass are naturally of the same order of magnitude, our model
  generically allows for light Higgs boson masses. Realistic mass
  ratios are related to particular RG trajectories with a
  ``walking'' mid-momentum regime.

\end{abstract}


\maketitle

\section{Introduction}
\label{sec:intro}

The chiral, gauge and flavor structures of the standard model lie at the heart
of its greatest successes. At the same time, they also mark its fundamental
deficits: the chiral structure requires a scalar Higgs field which suffers
from a severe triviality problem, suggesting that the conventional Higgs
sector is not a fundamental quantum field theory \cite{Wilson:1973jj}. Also
the product structure of the gauge symmetry involves a U(1) gauge symmetry
which has a similar (though less pressing) triviality problem \cite{Landau,%
  Gell-Mann:fq,Gockeler:1997dn,Gies:2004hy}. In addition, the perturbative
running of the Higgs sector appears to require unnaturally fine-tuned initial
conditions in order to separate the electroweak scale from the
Planck or GUT-like scales. In the same spirit, the diversity of scales in the
flavor sector so far has not found a convincing natural explanation.

Whereas many attempts to resolve these deficits are built on postulating new
degrees of freedom, new symmetries or new quantization rules, we wish to take
a fresh look at conventional systems within quantum field theory, relying on
the degrees of freedom already observed in experiments. In the present work,
we even consider a reduced model involving a chiral Higgs-Yukawa system with a
non-abelian gauged left-handed sector, which can be viewed as the scalar
Higgs-sector chirally coupled to a top-bottom fermion sector and a
left-handed SU($\NL$) gauge group. Many aspects of the weak-coupling
behavior of this model are straightforwardly accessible in perturbation
theory, showing some indications of the structural deficits mentioned above for the
standard model, in particular, the triviality and the hierarchy problem of the 
Higgs sector. 

In the present work, we therefore explore the model using the
functional renormalization group (RG) as a nonperturbative
method. Whereas all perturbative physics still remains included, we
can specifically address non-perturbative features to which naive
perturbation theory is completely blind. In fact, one such potentially
important feature in this context are threshold phenomena as has been
pointed out in \cite{Gies:2009hq, Gies:2009sv}. Threshold effects
such as the decoupling of massive modes is non-perturbative in the
sense that dynamically generated masses can be proportional to the
coupling. A decoupling that proceeds, for instance, inversely
proportional to some power of the mass therefore cannot have a naive
perturbative Taylor expansion in powers of the coupling without
spoiling the physical threshold behavior at any finite order of this
expansion. It is important to emphasize that this statement holds
independently of the coupling strength. In fact, the threshold
phenomena relevant for the present work turn out to be active in the
weak-coupling region of the model. In particular, the gauge
interactions are fully in the domain of asymptotic freedom. Of
  course, methods to deal with threshold phenomena have also been
  developed within perturbative approaches \cite{Wetzel:1981qg} and
  are commonly used to follow the RG evolution in the standard model
  \cite{Schroder:2005hy}. Hence, we expect that our results should
  also be reproducible within a perturbative treatment that properly
  accounts for threshold effects, for instance, in a mass-dependent RG
  scheme.

The scenario developed in the present work builds on the concept of
asymptotic safety \cite{Weinberg:1976xy} which is a generalization of
asymptotic freedom to non-Gau\ss ian, i.e. interacting, UV fixed
points. A quantum field theory can be UV complete if its RG trajectory
approaches a fixed point in the UV, such that a UV cutoff $\Lambda$
can be sent to infinity, $\Lambda\to \infty$. Though standard
perturbative renormalizability is included in this scenario if the
fixed point occurs at zero coupling, it is not mandatory for UV
completeness, but is merely a criterion for the applicability of
perturbation theory. Asymptotically safe theories are well known and
understood in lower-dimensional fermionic systems
\cite{Rosenstein:pt,Braun:2010tt,Parisi:1975im,Chandrasekharan:2011mn}. Most prominently,
asymptotic safety has by now become an established scenario for a UV
complete quantum theory of gravity \cite{Reuter:1996cp}. In this
  larger context of asymptotically safe gravity, new fixed point
  structures can also arise in the combined gravity-scalar
  \cite{Percacci:2003jz}, gravity-Yukawa \cite{Zanusso:2009bs},
  gravity-fermion \cite{Eichhorn:2011pc}, or gravity-photon
  \cite{Harst:2011zx} sector, potentially curring the UV problems of
  the standard model.

The present work continues the search for asymptotically safe Yukawa
models initiated in \cite{Gies:2009hq, Gies:2009sv}. However, our new
findings including a gauge sector go beyond those scenarios, as the
near-conformal behavior required for a non-Gau\ss ian fixed point does
not occur in the couplings and the vacuum expectation value, but is
shifted to the mass parameters of the model in the Higgs regime. This
gives rise to a novel and unconventional asymptotic safety scenario
for gauged Higgs-Yukawa systems which could be active in the
electroweak sector of the standard model.  If so, such a
  scenario not only modifies the UV behavior of the standard model
  potentially curing some deficits, but may also have
  implications for the infrared behavior. For instance, the accessible
  space of mass and coupling parameters can be constrained as a result
  of the RG flow. First results along this line will be discussed
  below.

The article is organized as follows. In Sect.~\ref{sec:model}, we
motivate the model in the context of earlier work, emphasizing the
fact that the construction of an asymptotically safe scenario for the
standard model Higgs sector appears to favor a chiral gauge
structure. Section \ref{sec:RG} details the application of the
functional RG technique to the present problem and summarizes our
results for the RG flow equations evaluated to next-to-leading order
in a derivative expansion of the Higgs-Yukawa system. The fixed point
structure of the model facilitating an asymptotic safety scenario is
analyzed in Sect.~\ref{sec:FP}, revealing a line of fixed points with
suitable UV properties. We explicitly verify in Sect.~\ref{sec:six}
that the UV fixed points can be connected with the Higgs phase of the
model, such that the IR properties are qualitatively reminiscent to
those of the standard model. This demonstrates that the present chiral
gauged Higgs-Yukawa model can be a UV complete quantum field theory.
Conclusions are presented in Sect.~\ref{sec:conc}, and many important
details of the calculations are deferred to the appendices.

\section{Motivation of the model}
\label{sec:model}
 
The chiral gauged Higgs-Yukawa model investigated in the present work
can, of course, straightforwardly be motivated from the
(experimentally observed) Higgs sector of the standard model. However,
it is instructive to realize that also purely theoretical arguments
for the construction of an asymptotically safe Higgs-Yukawa system
lead to the same model in a natural way. 

Inspired by the asymptotic safety of fermionic models such as the
simple Gross-Neveu model in $2<d<4$ dimensions
\cite{Rosenstein:pt,Braun:2010tt}, it is tempting to
ask, whether this nonperturbative renormalizability at a non-Gau\ss
ian fixed point can also be extended to $d=4$ dimensions
\cite{Gies:2003dp,Bazzocchi:2011vr,Schwindt:2008gj}. Even though final
answers in complex models have not been given so far, simple toy
models such as the Gross-Neveu model with a discrete $\mathds{Z}_2$
symmetry show a vanishing of the required fixed points in the limit
$d\to 4$ (more precisely: the non-Gau\ss ian fixed point typically
merges with the Gau\ss ian fixed point for $d\to 4$)
\cite{Braun:2010tt} as is also suggested by lattice simulations
\cite{Kim:1994pg}.

Moreover, since purely fermionic models and Yukawa models with similar
mass generating phase transitions are typically in the same
universality class \cite{ZinnJustin:1991yn,Hasenfratz:1991it}, it is
reasonable to treat the fermionic and the bosonic degrees of freedom
on the same footing, i.e., both as fundamental. 

An extensive exploration of the simple $\mathds{Z}_2$-invariant Yukawa
model using the functional RG \cite{Gies:2009hq} has provided evidence
that (i) no non-Gau\ss ian fixed point exists in the symmetric regime,
whereas (ii) the symmetry-broken regime can give rise to suitable
fixed points, provided the symmetry-breaking condensate approaches a
fixed point and behaves nearly conformal. If this conformal
condensate behavior sets in, a threshold behavior is induced that can
naturally lead to a balancing of interactions up to the highest
scales. More concretely, the flow of the scalar vacuum
expectation value (vev) $\bar{v}$ can be parametrized by the dimensionless
combination  $\kappa={v}^2/(2k^2)$, the $\beta$ function of which has
the generic structure
\begin{equation}
\pat \kappa \equiv \pat \frac{{v}^2}{2 k^2} = -2 \kappa +\, \text{boson
  fluct.}-\, \text{fermion fluct.},
\label{eq:conformalvev}
\end{equation}
where $\pat = k\frac{d}{dk}$ and $k$ denotes an RG scale. In order to
induce a non-Gau\ss ian fixed point at positive $\kappa_\ast>0$,
facilitating a (near) conformal condensate behavior $v\sim k$,
the bosons obviously have to dominate over the fermion fluctuations
due to the relative minus sign. Whether or not this is the case
essentially depends on the number of degrees of freedom of the
model. The analysis of \cite{Gies:2009hq} containing one real scalar
field and $\Nf$ Dirac fermions revealed that the necessary bosonic
dominance occurs only for an unphysical value of $\Nf\lesssim 0.3$.

An elegant way to enhance the boson fluctuations (without unnaturally
increasing the number of boson fields) was identified in
\cite{Gies:2009sv}: chiral Yukawa couplings of $\NL$ complex scalar
fields $\phi^a$ with $\NL$ left-handed fermions $\psi^a_{\text{L}}$
and a single right-handed fermion $\psi_{\text{R}}$ leads to an $\sim
\NL$ enhancement of the boson fluctuations whereas the fermion
fluctuations remain of $\sim \mathcal{O}(1)$. This is already a first indication that
asymptotically safe scenarios prefer chiral Yukawa systems. In
\cite{Gies:2009sv}, suitable non-Gau\ss ian fixed points where
discovered for a wide range of $\NL$ including $\NL=2$ in a
leading-order derivative expansion analysis.  Moreover, one of the
admissible fixed points has only one UV-attractive direction, thus
implying that only one physical parameter has to be fixed, e.g., the
vev $v=246$GeV, whereas all other IR quantities such as the Higgs or
the top mass would be a pure prediction of the theory.

However, these suitable fixed points apparently get destabilized at
next-to-leading order in the derivative expansion \cite{Gies:2009sv}
for a physical reason: the derivative expansion assumes that field 
amplitudes remain sufficiently slowly varying at a given RG scale
during the flow. But in the chiral Yukawa model, massless Goldstone
modes occur in the broken regime which together with the massless
bottom-type fermions of the model induce strong contributions which
are not damped by threshold effects. We expect this argument to be
rather generic: non-Gau\ss ian threshold-induced fixed points in
Yukawa models with continuous symmetries are likely to be destabilized
by massless Goldstone modes in the broken regime. 

There is one particular mechanism to avoid massless Goldstone modes in
systems with broken continuous symmetries: the Higgs mechanism
\cite{Englert:1964et}. Hence, the search for non-Gau\ss ian fixed
points in Yukawa systems naturally leads us to the inclusion of a
chiral gauge sector in our model. In addition to ``eating up'' the
scalar Goldstone modes, we expect the gauge bosons to also contribute
in a beneficial way to the stabilization of the scalar condensate,
cf. \Eqref{eq:conformalvev}.

On the other hand, already at this point, we can expect that the
picture of the conformal threshold behavior might be modified upon the
inclusion of gauge fields. The reason is that non-abelian gauge
theories are asymptotically free. Whereas for any finite value of the
gauge coupling $g^2$, we may find stable non-Gau\ss ian fixed points
along the lines of \cite{Gies:2009hq,Gies:2009sv}, the UV limit where
$g^2 \to 0$ must ultimately leave its imprints also in the Yukawa
sector. It is one of the main results of the present work
that the conformal threshold behavior indeed persists, however not in
the form of near-conformal condensate and couplings, but in the form
of near-conformal mass parameters. 

The reasoning of this section leads us to consider chiral gauged
Higgs-Yukawa models with a standard classical action of the form (here
and in the following, we work in Euclidean space)
\begin{align}\label{eq:classicalaction}
S_\mathrm{cl}= &\int\!\! d^dx\Big[ \frac{1}{4}F_{\mu \nu}^iF^{i\mu \nu}+(D^{\mu}\phi)^{\dagger}(D_{\mu}\phi)+\bar{m}^2\rho+\frac{\bar{\lambda}}{2}\rho^2\nonumber\\
&+i(\bar{\psi}_\mathrm L^a\slashed{D}^{ab}\psi_\mathrm L^b+\bar{\psi}_\mathrm R\slashed{\partial}\psi_\mathrm R)
+\bar{h}\bar{\psi}_\mathrm R\phi^{\dagger a}\psi_\mathrm L^a-\bar{h}\bar{\psi}_\mathrm L^a\phi^{a}\psi_\mathrm R\Big]
\end{align}
where $\rho:=\phi^{a\dagger}\phi^a$. The classical parameter space is
spanned by the boson mass $\bar m$, the scalar self-interaction
$\bar\lambda$, the Yukawa coupling $\bar h$ and the gauge coupling
$\bar{g}$ which occurs in the covariant derivatives for the matter
fields in the fundamental representation of the gauge group ($a,b,\dots =
1, \dots, \NL$),
\be
D_\nu^{ab}=\partial_\nu\delta^{ab}-i\bar{g} W_\nu^i (T^i)^{ab},
\label{eq:covder}
\ee
where $W_\nu^i$ denotes the Yang-Mills vector potential. The
fermionic field content consists of a left-handed $\NL$-plet
  (e.g., a top-bottom doublet for SU($\NL=2$)) and one right-handed
  fermion (e.g., the right-handed top-quark component); right-handed
  bottom-type components are not considered, such that only the top
  quark can become massive upon symmetry breaking. The fermions are
considered to occur in $\Ng$ generations; the generation structure
does not have any nontrivial interplay with the gauge or scalar
sector, such that the corresponding generation index is
suppressed. The general calculations of the present work in principle
hold for any simple Lie group, hence we keep the notation general. In
concrete calculations we will confine ourselves to SU($\NL=2$). We
expect the mechanisms presented below to hold for any gauge
group.\footnote{ Of course, the present model has perturbative gauge
  anomalies for SU($\NL\geq 3$) \cite{Ball:1988xg}. For SU($\NL=2$) or
  SP($\NL$), the model has a global Witten anomaly for odd $\Ng$
  \cite{Witten:1982fp}. In these anomalous cases, the model cannot be
  a consistent quantum field theory as it stands. The RG flows
  determined below should in such cases be viewed as a projection of a
  larger (unspecified) anomaly-free model, e.g., the standard model
  with only one generation, onto an effectively reduced theory
  subspace.~\label{foot:anomaly}} The generators of the gauge group
satisfy the corresponding algebra, $[T^i,T^j]=if^{ijk}T^k$ with
structure constants $f^{ijk}$, and the nonabelian field strength in
\Eqref{eq:classicalaction} is given by $F_{\mu \nu}^i=\partial_\mu
W_\nu^i-\partial_\mu W_\nu^i+\bar{g} f^{ijl}W^j_\mu W^l_\nu$, where
$i,j,k, \dots$ denote adjoint indices. For the scalar potential, we
only consider the invariant $\rho:=\phi^{a\dagger}\phi^a$. For later
convenience, we remark that the complex scalar field can equally well
be expressed in terms of $2\NL$ real scalar fields,
\begin{eqnarray}
 \phi^a = \frac{1}{\sqrt{2}}(\phi_1^a + i\phi_2^a),\quad
 \phi^{a\dagger} = \frac{1}{\sqrt{2}}(\phi_1^a - i\phi_2^a)\,,
\end{eqnarray}
where $\phi_1^a, \phi_2^a\in\mathbb{R}$. In addition to the local
gauge symmetry, the model is invariant under a global
U(1)$_{\text{L}}\times$U(1)$_{\text{R}}$ symmetry, where the fermions
transform under their corresponding chiral component and the scalar
transforms under both with opposite charges. If the gauge
symmetry is chosen to be SU($\NL$), its global part and the
U(1)$_{\text{L}}$ are subgroups of a global U($\NL$)$_{\text{L}}$
symmetry.  In the following, we will analyze the RG flow of generic
effective actions in a theory space inspired by the classical action
given in \Eqref{eq:classicalaction} and its symmetries.

\section{RG flow of the model}
\label{sec:RG}

\subsection{Functional RG}
In the present work, we study the chiral gauged Higgs-Yukawa model
using the functional RG. More precisely, we study the RG flow of
effective action functionals $\Gamma_k$ that are spanned by the same
field content and the same symmetries as \Eqref{eq:classicalaction}.
Here, the scale $k$ denotes an IR cutoff parametrizing those
fluctuations with momenta $p^2\lesssim k^2$ that still have to be
fully integrated out to arrive at the full effective action
$\Gamma=\Gamma_{k\to0}$. The latter corresponds to the standard
generating functional of 1PI correlation functions encoding the
physical properties of the theory. 

The set of functionals $\Gamma_k$ hence defines a one-parameter family
of effective actions that relate the physical long-range behavior for
$k\to 0$ with a microscopic action functional at $k\to \Lambda$, where
$\Lambda$ denotes a microscopic UV scale, such that $\Gamma_\Lambda$
is related to the ``classical action to be quantized''. The trajectory
interconnecting all these scales is determined by the Wetterich
equation \cite{Wetterich:1992yh}
\begin{equation}\label{flowequation}
	\partial_t\Gamma_k[\Phi]
        =\frac{1}{2}\mathrm{STr}\{[\Gamma^{(2)}_k[\Phi]+R_k]^{-1}(\partial_tR_k)\}\, .
\end{equation}
Here $\Gamma^{(2)}_k$ is the Hessian, i.e., the second functional
derivative with respect to the field $\Phi$, representing a collective
field variable for all bosonic or fermionic degrees of freedom. The
momentum-dependent regulator function $R_k$ encodes the suppression of
IR modes below a momentum scale $k$, for reviews see \cite{ReviewRGgauge,ReviewRG}.

Whereas \Eqref{flowequation} in conventional applications is solved
subject to an initial condition $S_\Lambda \simeq S_{\text{cl}}$ (the
bare action), we use the flow equation also to search for suitable
initial conditions in the vicinity of UV attractive fixed points. If
such fixed points exist, trajectories can be constructed that are UV
complete by approaching and ultimately hitting the fixed point in the limit
$\Lambda\to \infty$. The corresponding system together with
its flow to the IR represents a quantum field theory which can be
valid on all scales. 

For the present work, the crucial property of the functional RG
evolution is the fact that the computation of correlation functions
involves the exact (regularized) propagator at a scale $k$ given by
$[\Gamma^{(2)}_k[\Phi]+R_k]^{-1}$ in \Eqref{flowequation}. Especially,
if a dynamically generated mass exists at some scale $k$, it is
included as a corresponding gap in the self-energy. By contrast, naive
perturbation theory consists of an expansion about zero coupling which
is blind to dynamically generated masses. Of course, the latter can be
included in a reorganized perturbative expansion, but the
functional RG does so in a self-consistent and RG-improved manner. In
this way, we can particularly well deal with the threshold regime
where the mass generation sets in dynamically.

As we are dealing with a gauge theory, the RG flow has to be
constructed such that gauge symmetry is preserved. As in standard
continuum calculations, gauge fixing is required such that gauge
symmetry is encoded in constraints (generalized Ward
identities). While it is by now well understood how to deal with this
issue in the non-perturbative strong-coupling domain
\cite{Reuter:1993kw,Ellwanger:1994iz,Pawlowski:2001df,Gies:2002af,Branchina:2003ek,ReviewRGgauge},
the present work only requires the weak-coupling limit of the gauge
sector essentially on a one-loop level. For this, we will use the
standard background field formalism \cite{Abbott:1980hw}; details of
this part of the calculation can be found in App.~\ref{sec:appgauge}.

In the following, we use the $R_\alpha$ gauge (or its background-field
variant for the computation of the gauge sector, see
App.~\ref{sec:appgauge}). For its definition in the present context,
let us first decompose the scalar field into the bare vev $\bar{v}$ and
the fluctuations $\Delta\phi$ about the vev
\begin{equation}\label{fluctandvev}
\phi^a=\frac{\bar{v}}{\sqrt{2}}\hat{n}^a+\Delta\phi^a, \quad
\Delta\phi^a=\frac{1}{\sqrt{2}}(\Delta\phi_1^a + i\Delta\phi_2^a),
\end{equation}
where $\hat{n}$ is a unit vector ($\hat{n}^{\dagger}_a\hat{n}^a=1$)
defining the direction of the vev in fundamental Yang-Mills
space. Then, the gauge fixing condition is given by
\begin{equation}\label{eq:gaugefixingcond}
G^i(W)=\partial_\mu W^i_\mu + i\alpha
\bar{v}\bar{g}(T^i_{{\hat{n}}\check{a}}\Delta\phi_1^{\check{a}}+iT^i_{{\hat{n}}a}\Delta\phi_2^a)=0, \quad \check{a}\neq\hat{n},
\end{equation}
where $\alpha$ is a gauge-fixing parameter interpolating between the
unitary gauge at $\alpha\to\infty$ and the Landau gauge at
$\alpha\to 0$. Here and in the following, the label ${\hat{n}}$ in
place of a fundamental color index denotes the contraction of that
index with the unit vector ${\hat{n}}$ (or ${\hat{n}}^\dagger$,
depending on the position of the index). In \Eqref{eq:gaugefixingcond} the component 
$\Delta \phi_1^{\hat{n}}$ is not included in the sum over $\check{a}$.
This implies that the gauge fixing only involves the Goldstone-boson
directions and not the radial mode. The gauge fixing is implemented by
including a gauge-fixing term in the action
\begin{eqnarray}\label{eq:gaugefixing}
S_\mathrm{gf}&=&\frac{1}{2\alpha}\int\!\! d^dx\, G^i(W)G^i(W),\nonumber
\end{eqnarray}
as well as a Faddeev-Popov term localized in terms of ghost fields
$c^i$ and $\bar{c}^i$ with a bare action
\be
S_\mathrm{gh}= \int\!\! d^dx\, \bar{c}^i\mathcal{M}^{ij}c^j.\nonumber
\ee
The Faddeev-Popov operator is given by
\begin{equation}\label{eq:FadeevPopovoperator}
\mathcal{M}^{ij}=-\partial^2\delta^{ij}-\bar{g}f^{ilj}\partial_{\mu}W^{l\mu}
+\sqrt{2}\alpha \bar{v}\bar{g}^2T_{{\hat{n}}\check{a}}^iT^j_{\check{a}b}\Delta\phi^b,
\end{equation}
again excluding $a=\hat{n}$ in the sum over $\check{a}$.

With these preparations, we can now write down the space of action
functionals considered in this work:
\begin{align}\label{eq:truncation}
\Gamma_k = \int d^dx &\Big[ U(\rho)+Z_{\phi}(D^{\mu}\phi)^{\dagger}(D_{\mu}\phi)\\
&+i(Z_{\mathrm L}\bar{\psi}_\mathrm L^a\slashed{D}^{ab}\psi_\mathrm L^b
+Z_{\mathrm R}\bar{\psi}_\mathrm R\slashed{\partial}\psi_\mathrm R)\nonumber\\
&+\bar{h}\bar{\psi}_\mathrm R\phi^{a\dagger}\psi_\mathrm L^a
-\bar{h}\bar{\psi}_\mathrm L^a\phi^{a}\psi_\mathrm R\nonumber\\
&+\frac{Z_{W}}{4}F_{\mu \nu}^iF^{i\mu \nu} +\frac{Z_{\phi }}{2\alpha} G^iG^i
-\bar{c}^i\mathcal{M}^{ij}c^j \Big]\, .\nonumber
\end{align}
All couplings, wave function renormalizations
$Z_{\phi,\text{L},\text{R},W}$ and the effective scalar potential
$U(\rho)$ are taken to be $k$ dependent. The scalar sector corresponds
to a next-to-leading order derivative expansion of the action. In
addition, the flows of the Yukawa coupling, the gauge coupling and the
wave function renormalizations are evaluated in the presence of a
$k$-dependent minimum of the potential in order to properly account
for the threshold phenomena.  

As \Eqref{eq:truncation} already indicates, 
we ignore any nontrivial running of the ghost sector and of the gauge parameter $\alpha$
and drop any higher order gauge-field operators, as this is not necessary for
an exact flow of the gauge coupling at one-loop order.
For the actual computation of the latter using the background field method, the
ordinary derivatives in the gauge fixing and ghost terms are replaced
by covariant derivatives $\bar{D}$ w.r.t. the background field
$\bar{W}$, cf. App.~\ref{sec:appgauge}. The Yukawa sector remains
however unaffected by the background field.

To sum up, the subset of theory space we are considering is
parametrized by $Z_{\phi }$, $Z_{\mathrm L }$, $Z_{\mathrm R }$,
$Z_{W}$, ${\bar h}$, $\bar{v}$ and all the parameters contained in
$U$ different from $\bar{v}$ itself. (The running of the gauge coupling
$\bar{g}$ in the background field method is related to the wave
function renormalization $Z_{W}$, see below).

It is also useful to introduce a simplifying notation for the masses
in the symmetry-broken regime, which are directly related to the parameters listed
above. The (unrenormalized) mass matrix for the gauge bosons is given
by
\be\label{gaugemass}
\bar{m}^{2\ \ ij}_W=\frac{1}{2}Z_\phi \bar{g}^2 \bar{v}^2 \{T^i,T^j\}_{{\hat{n}}{\hat{n}}}.
\ee
Since it is diagonalizable, we can choose a basis in adjoint color
space where
\be\label{gaugemassdiag}
\bar{m}^{2\ \ ij}_W=\bar{m}^2_{W,i}\delta^{ij}\quad\text{(no sum over $i$)}\, .
\ee
The scalar mass matrix reads
\be
\bar{m}_\phi^{2\,ab}=\bar{v}^2U''\left(\frac{\bar{v}^2}{2}\right)\hat{n}^a\hat{n}^{\dagger b}\, .\nonumber
\ee
In a diagonalizing basis, we have
$\bar{m}_\phi^{2\,ab}=\bar{m}_{\phi,a}^2\delta^{ab}$ (no sum over
$a$), with, of course, vanishing eigenvalues for the would-be
Goldstone modes corresponding to the broken generators in this
gauge.  Furthermore the (unrenormalized) ``top mass'' i.e. the mass
of the $\psi^{\hat n}$ mode, is given by
\be\label{eq:topmass}
\bar{m}_\text{t}=\frac{\bar{h}\bar{v}}{\sqrt{2}}\, .  
\ee
The corresponding renormalized quantities include appropriate factors
of the wave function renormalizations, see below.  {Incidentally,
  the above reasoning for identifying the particle spectrum follows
  that used in straightforward perturbative considerations. We
  emphasize that a proper gauge- and scheme-independent identification
  requires careful nonperturbative considerations, see e.g.
  \cite{Frohlich:1981yi,Maas:2012tj}.}

\subsection{Dimensionless variables}

Whereas the long-range observables are dimensionful quantities
expressed, for instance, in terms of an absolute measurement scale, the
search for UV fixed points requires dimensionless variables. If the
system approaches a conformal behavior, it is expected to become
self-similar, i.e., to look the same independently of the measurement
scale. Also the flow equations can conveniently be expressed in terms
of dimensionless renormalized variables. For this, we define the
corresponding Yukawa and gauge couplings,
\begin{equation}
h^2=\frac{k^{d-4}\bar{h}^2}{Z_{\phi}Z_{\mathrm{L}}Z_{\mathrm{R}}},\quad
g^2=\frac{\bar{g}^2}{Z_Wk^{4-d}}, \label{eq:dimlesshg}
\end{equation}
as well as the dimensionless effective potential
\begin{equation}
u(\tilde{\rho})=k^{-d}U(Z_{\phi}^{-1}k^{d-2}\tilde{\rho}),\quad
\tilde{\rho}=\frac{Z_\phi \rho}{k^{d-2}}. \label{eq:dimlessu}  
\end{equation}
The effective potential is expressed in terms of the dimensionless
renormalized field variable $\tilde{\rho}$. If the system is in the
spontaneously symmetry-broken (SSB) regime, we use the dimensionless
renormalized minimum of the potential,
\begin{equation}
\kappa=\frac{Z_\phi
  \bar{v}^2}{2k^{d-2}}=\tilde{\rho}_{\min}. \label{eq:kappa1}
\end{equation}
Correspondingly, an expansion of the effective potential about the
minimum can be useful in the SSB regime,
\begin{eqnarray}\label{eq:uexpSSB}
 	u&=&\sum_{n=2}^{N_\text{p}}\frac{\lambda_{n}}{n!}(\tilde{\rho}-\kappa)^n\\ 
&=&\!\frac{\lambda_{2}}{2!}(\tilde{\rho}-\kappa)^2
        \!+\frac{\lambda_{3}}{3!}(\tilde{\rho}-\kappa)^3+\cdots\nonumber\,.
\end{eqnarray}
In the symmetric regime where $\tilde{\rho}_{\text{min}}=0$, we use
the expansion
\begin{eqnarray}\label{eq:symeffpot}
  u\!\!&=&\!\!\sum_{n=1}^{N_p}\frac{\lambda_{n}}{n!}\tilde{\rho}^n 
  = m^2\tilde{\rho}+\frac{\lambda_{2}}{2!}\tilde{\rho}^2
  +\frac{\lambda_{3}}{3!}\tilde{\rho}^3+\cdots
\end{eqnarray}
Note that the expansion coefficients $\lambda_{n}$ in
Eqs. \eqref{eq:uexpSSB} and \eqref{eq:symeffpot} are generally not
identical in the two different regimes; also their flow will be
different.  Still the standard $\phi^4$ coupling in both cases is
related to $\lambda_2$.  The contribution of the field
renormalizations deviating from canonical scaling is encoded in the
scale-dependent anomalous dimensions
\begin{eqnarray*}
	\eta_{\phi}&=&-\partial_t \log Z_{\phi}\, ,\qquad  \eta_W=-\partial_t\log Z_W \\
        \eta_{\mathrm L}&=&-\partial_t \log Z_{\mathrm L}\, ,\qquad
         \eta_{\mathrm R}=-\partial_t \log Z_{\mathrm R} \, .
\end{eqnarray*}
Setting the anomalous dimensions to zero defines the leading-order
derivative expansion. At next-to-leading order, it is important to
distinguish between the running of $Z_{\mathrm{L}}$ and that of
$Z_{\mathrm{R}}$ as they acquire different loop contributions, see
below.

As we compute the running of the gauge coupling with the
background-field method, we can make use of the fact that the product
of bare coupling and bare background gauge field $\bar g W_\mu^a$ is a
renormalization group invariant combination in the background-field
gauge. Therefore, the running of the gauge coupling is tightly linked
to that of the background-field renormalization \cite{Abbott:1980hw},
implying that
\begin{equation}
\beta_{g^2} = \pat g^2 = (d-4+ \eta_W) g^2. \label{eq:betagqd}
\end{equation}
For an analysis of the threshold behavior of the system, also the
dimensionless renormalized mass parameters turn out to be useful:
\begin{equation}
\mu^2_{W,i}=\frac{\bar{m}^2_{W,i}}{Z_W k^2},\quad \mu^2_{\phi,a}=\frac{\bar{m}^2_{\phi,a}}{Z_\phi k^2},
\quad \mu_{\text{t}}^2=\frac{\bar{m}_{\text{t}}^2}{Z_\mathrm L
    Z_\mathrm R k^2}\,. 
\label{eq:defmu}
\end{equation}
From a general perspective, the dimensionless renormalized formulation
can only implicitly depend on our RG scale $k$, as the latter is
dimensionful. Therefore, a fixed point associated with conformal
behavior corresponds to the vanishing of the $\beta$ functions of all
dimensionless variables. However, the above given list of
dimensionless couplings contains some redundancy, as, for instance,
the dimensionless top mass is related to the Yukawa coupling and the
potential minimum, and similarly for the gauge boson and Higgs
masses. A more precise statement thus is that a fixed point exists, if
the $\beta$ functions for a complete linearly independent set of
dimensionless variables vanish. For the moment, we keep this
redundancy of variables in order to determine a suitable choice of
variables below.

As we will specialize to the case of SU($\NL=2$) below, let us already
list here the relations between all couplings for this case:
\begin{equation}
\mu_W^2=\frac{1}{2} g^2 \kappa, \quad \mu_{\text{H}}^2=2
\lambda_2\kappa, \quad \mu_t^2=\kappa h^2. \label{eq:dimlessmuSU2}
\end{equation}
The dimensionful renormalized masses of the system can
straightforwardly be obtained by trivial multiplication with the
scale,
\begin{equation}
m_W^2=\mu_W^2 k^2, \quad m_{\text{H}}^2=\mu_{\text{H}}^2 k^2, \quad
m_{\text{t}}^2= \mu_t^2 k^2. \label{eq:massesSU2}
\end{equation}
Together with the dimensionful renormalized vev, 
\begin{equation}
v= \sqrt{ 2 \kappa}\, k^{(d-2)/2} \equiv Z_\phi^{1/2} \bar{v}, \label{eq:vev}
\end{equation}
this list constitutes our long-range observables to be computed from the flow
towards the IR.

\subsection{Flow equations for the matter couplings}

Large parts of the calculation of the flow equation in the matter
sector can be done in arbitrary Euclidean space dimension $d$, for any
number of left-handed fermion components $\NL$, number of
  generations $\Ng$ and any gauge group with a corresponding
dimension $\dG$ of its adjoint representation (e.g., $\dG=\NL^2-1$ for
SU($\NL$)). Furthermore, the dimension of the representation of the
Clifford algebra for the chiral fermions will be abbreviated by
$d_\gamma$. As a special case, we have chosen to work in Landau gauge
$\alpha\to 0$ (or Landau-DeWitt gauge for the background-field part)
for reasons of simplicity. Also, Landau gauge is known to be a fixed
point of the RG flow \cite{Ellwanger:1995qf} and hence is a
self-consistent choice.

Later on, we specialize to $d=4$, SU($\NL=2$) with $\dG=3$ and
$d_\gamma=2$. Also, we use the linear regulator that
is optimized for the present truncation \cite{Litim:2001up}. We
have not found any indication that the qualitative features
described below depend on any of these concrete choices. 

We defer the details of the calculation as well as the properties of
the threshold functions $l,m,a$ occurring below with various sub- and
superscripts and  parametrizing the decoupling of massive modes
to App.~\ref{section:threshold}. 

Let us start with the flow of the effective potential which is driven
by scalar (B superscript), Dirac fermion (F), left-handed Weyl fermion (L), as well as gauge boson (G)
and ghost (gh) fluctuations. Introducing the abbreviation $v_d=1/(2^{d+1}\pi^{d/2} \Gamma(d/2))$
the flow of the potential is described by
\begin{widetext}
\begin{eqnarray}\label{floweq:potential}
\partial_t u = -d u + (d-2 + \eta_\phi)\tilde\rho u' +  2 v_d \Big\{ 
- 2 \dG l_0^{(\mathrm{gh})d}\left( 0 \right) + \sum_{i=1}^{\dG} \left[ 
(d-1)l_{0\mathrm T}^{(\mathrm{G})d}\left(\mu_{W,i}^2 (\tilde\rho)\right) + l_{0\mathrm L}^{(\mathrm{G})d}\left( 0 \right) \right] \nonumber\\
+(2\NL-1)l_0^{(\mathrm{B})d}\left( u' \right) + l_0^{(\mathrm{B})d}\left(u' + 2 \tilde\rho u'' \right) - d_\gamma \Ng\left[ (\NL-1)l_{0}^{(\mathrm L)d}\left( 0 \right) + 2 l_{0}^{(\mathrm F)d}\left( \tilde\rho h^2 \right) \right] \Big\}.
\end{eqnarray}
\end{widetext}
where $\mu_{W,i}^2 (\tilde\rho)$ are defined as functions of the full scalar field in analogy with Eqs.~\eqref{gaugemass},\eqref{gaugemassdiag},
reducing to the dimensionless gauge boson renormalized masses for $\tilde\rho=\kappa$.
The full flow of $u$ \eqref{floweq:potential} can be used to extract
the flows of the coefficients of the potential expansions
Eqs.~\eqref{eq:uexpSSB},\eqref{eq:symeffpot} in both regimes. For the
flow of the minimum $\kappa$ in the SSB regime, we use the fact that
the first derivative of $u$ vanishes at the minimum,
$u'(\kappa)=0$. This implies
\begin{eqnarray}
 0= \pat u'(\kappa)&=&\partial_t u'(\tilde\rho)|_{\tilde\rho=\kappa}
 +(\partial_t \kappa)u''(\kappa)\nonumber\\
 \Rightarrow \partial_t \kappa&=&-\frac{1}{u''(\kappa)}\partial_t
 u'(\tilde\rho)|_{\tilde\rho=\kappa}\,. \label{eq:kappa}
\end{eqnarray}
Whereas the flow in the symmetric regime is unambiguous, a subtlety
arises in the SSB regime: here, the flow of the Yukawa coupling and
the scalar anomalous dimension for the would-be Goldstone modes can,
in principle, be different from those of the radial mode. As the
Goldstone modes as such are not present in the standard model, we
compute the Yukawa coupling and the scalar anomalous dimension by
projecting the flow onto the radial scalar operators in the SSB
regime. Note that this strategy is different from that used for
critical phenomena in other Yukawa or bosonic systems, where the
Goldstone modes typically dominate criticality.  Accordingly, the flow
of the Yukawa coupling $h$ can be derived and we find the same
result already presented in \cite{Gies:2009sv}, that is
\begin{widetext}
\begin{eqnarray}\label{floweq:yukawa}
\partial_t h^2 &=&(d-4+\eta_{\phi}+\eta_{\mathrm{L}}+\eta_{\mathrm{R}})h^2+ 4v_d h^4\Big\{
(2\tilde{\rho}u'') l_{1,2}^{(\mathrm{FB})d}(\tilde{\rho}h^2, u')-(6\tilde{\rho}u''
+4\tilde{\rho}^2u''')l_{1,2}^{(\mathrm{FB})d}(\tilde{\rho}h^2, u'+2\tilde{\rho}u'')\nonumber\\
&-&l_{1,1}^{(\mathrm{FB})d}(\tilde{\rho}h^2, u')+l_{1,1}^{(\mathrm{FB})d}(\tilde{\rho}h^2, u'+2\tilde{\rho}u'')
+(2\tilde{\rho}h^2)l_{2,1}^{(\mathrm{FB})d}(\tilde{\rho}h^2, u')
-(2\tilde{\rho}h^2)l_{2,1}^{(\mathrm{FB})d}(\tilde{\rho}h^2,
u'+2\tilde{\rho}u'')\Big\}_{\tilde{\rho}=\tilde{\rho}_{\text{min}}}
\,. 
\end{eqnarray}
Note that the whole expression in curly braces vanishes in the
symmetric regime where $\tilde{\rho}=0$. Also, we observe that no gauge
contributions to the running of this coupling occur which is a special feature of
Landau gauge; a brief explanation of this fact is given in
App.~\ref{sec:deriv-matter-flow}. Finally, the anomalous dimensions read
\begin{eqnarray}\label{floweq:etaphi}
\eta_{\phi}&=&\frac{8 v_d}{d}\Big\{
\tilde\rho(3 u'' +2\tilde\rho u''')^2 m_{2,2}^{(\mathrm{B})d}(u'+2\tilde\rho u'',u'+2\tilde\rho u'')
+(2N_{\mathrm{L}}-1)\tilde\rho u''^2 m_{2,2}^{(\mathrm{B})d}(u',u')\nonumber\\
&+&d_\gamma \Ng h^2\left[ m_4^{(\mathrm{F})d}(\tilde\rho h^2)-\tilde\rho h^2 m_2^{(\mathrm{F})d}(\tilde\rho h^2)\right]\Big\}\nonumber \\
&+&\frac{8 v_d (d-1)}{d} \Bigg\{-2g^2 \sum_{a=1}^{\NL}\sum_{i=1}^{\dG} T^i_{{\hat n}a}T^i_{a{\hat n}} \, 
l_{1,1}^{(\mathrm{BG})d}\left( u', \mu_{W,i}^2 \right) 
+\sum_{i=1}^{\dG} \frac{\mu_{W,i}^4}{\tilde\rho} \left[ 
2 a_1^d\left(\mu_{W,i}^2 \right) + m_2^{(\mathrm{G})d}\left(\mu_{W,i}^2 \right) \right]\Bigg\}\Bigg|_{\tilde{\rho}=\tilde{\rho}_{\text{min}}}
\end{eqnarray}
\begin{equation}\label{floweq:etaR}
\eta_\mathrm R = \frac{4v_d}{d}h^2 \left[ m_{1,2}^{(\mathrm{LB})d}(\tilde\rho h^2, u' + 2\tilde\rho u'') 
+ m_{1,2}^{(\mathrm{LB})d}(\tilde\rho h^2, u') + 2(\NL-1)m_{1,2}^{(\mathrm{LB})d}(0, u') \right]_{\tilde{\rho}=\tilde{\rho}_{\text{min}}}
\end{equation}
\begin{eqnarray}\label{floweq:etaLgeneral}
\eta_\mathrm L &=& \frac{4v_d}{d} h^2 \left[ m_{1,2}^{(\mathrm{RB})d}(\tilde\rho h^2, u' + 2\tilde\rho u'') 
+ m_{1,2}^{(\mathrm{RB})d}(\tilde\rho h^2, u') \right] + \frac{8 v_d(d-1)}{d}g^2 \Bigg\{\nonumber\\ 
&&\sum_{i=1}^{\dG} (T^i_{{\hat n}{\hat n}})^2 \left[ 
m_{1,2}^{(\mathrm{LG})d}\left( \tilde \rho h^2, \mu_{W,i}^2 \right)
-m_{1,2}^{(\mathrm{LG})d}\left(0, \mu_{W,i}^2 \right)
- a_3^d\left( \tilde\rho h^2, \mu_{W,i}^2 \right)
+ a_3^d\left( 0, \mu_{W,i}^2 \right) \right]\nonumber\\
&&+\sum_{a=1}^{\NL}\sum_{i=1}^{\dG} T^i_{{\hat n}a}T^i_{a{\hat n}} \left[ 
m_{1,2}^{(\mathrm{LG})d}\left(0, \mu_{W,i}^2 \right) 
- a_3^d\left(0, \mu_{W,i}^2 \right) \right]\Bigg\}\Bigg|_{\tilde{\rho}=\tilde{\rho}_{\text{min}}}\,.
\end{eqnarray}
If the direction of the vev ${\hat n}$ has a single nonvanishing component in
the chosen basis of fundamental color algebra, i.e. if ${\hat n}^a\propto
\delta^{aA}$, the anomalous dimension of the left-handed fermion takes a
simpler form
\begin{eqnarray}\label{floweq:etaLsimple}
\eta_\mathrm L = &&\frac{4v_d}{d} h^2 \left[ 
m_{1,2}^{(\mathrm{RB})d}(\tilde\rho h^2, u' + 2\tilde\rho u'') 
+ m_{1,2}^{(\mathrm{RB})d}(\tilde\rho h^2, u') \right] \nonumber\\
&& + \frac{8 v_d(d-1)}{d}g^2 \sum_{a=1}^{\NL}\sum_{i=1}^{\dG} T^i_{Aa}T^i_{aA} \left[ 
m_{1,2}^{(\mathrm{LG})d}\left( \delta^{aA} \tilde \rho h^2, \mu_{W,i}^2 \right) 
- a_3^d\left( \delta^{aA} \tilde\rho h^2, \mu_{W,i}^2 \right) \right]\Bigg|_{\tilde{\rho}=\tilde{\rho}_{\text{min}}}\,.
\end{eqnarray}
The explicit form of these equations for the linear regulator can be found in App.~\ref{sec:deriv-matter-flow}.
\end{widetext}

\subsection{Flow equation for the gauge coupling}

Next, we list our results for the flow of the gauge coupling. As only the
weak-coupling flow of the gauge coupling is required for the present scenario,
we will be satisfied essentially with a one-loop approximation. For
consistency with the matter sectors, we pay special attention to the threshold
behavior, as worked out in greater detail in App.~\ref{sec:appgauge}. Since
the threshold behavior of the gauge sector depends a bit stronger on the gauge
group, we concentrate on the special case SU($\NL=2$) in $d=4$ Euclidean
dimensions. Following \Eqref{eq:betagqd}, the running gauge coupling can be
extracted from the wave function renormalization of the gauge field,
\begin{eqnarray}\label{floweq:gauge}
  \partial_t g^2 =\!\!&g^2&\!\! \eta_W,\\
  \eta_W =\frac{-g^2}{48 \pi^2}\Big(22\NL\,
  \mathrm L_W(\mu_{W,i}^2)\!\!\!&-&\!\!\!
  d_\gamma \Ng \mathrm L_\psi(\mu_{\text{t}}^2)
  -\mathrm L_\phi(\mu_{\phi,a}^2)\Big),\nonumber
\end{eqnarray}
where the form of the anomalous dimension follows solely from the fact that an
additive decomposition into gauge, fermion, and scalar loops can be performed
at one-loop order. The threshold functions $L_{W,\psi,\phi}$ parametrize the
decoupling of massive modes in the SSB regime. They are normalized by the
condition $L_{W,\psi,\phi}(0)=1$, such that the standard one-loop $\beta$
function for the gauge coupling is reobtained for massless fluctuations. We
also neglect here possible RG improvement from the dependence of these
threshold functions on the matter-field anomalous dimensions which contribute
at two-loop order. 

For SU$(2)$, we use in the present work
\begin{align}\label{floweq:gaugeLSU2}
\mathrm L_W(\mu_{W}^2)&=\frac{1}{44}\left(21+\frac{21}{1+\mu_{W}^2}+2\right)\, ,& \mu_{W}^2&=\frac{g^2\kappa}{2},\nonumber\\
\mathrm L_\psi(\mu_{\text{t}}^2)&=\frac{1}{2}\left(1+\frac{1}{1+\mu_{\text{t}}^2}\right)\, ,& \mu_{\text{t}}^2&=h^2\kappa,\nonumber\\
\mathrm L_\phi(\mu_{\text{H}}^2)&=\frac{1}{2}\left(1+\frac{1}{1+\mu_{\text{H}}^2}\right)\, ,& \mu_{\text{H}}^2&=2\lambda_{2}\kappa,
\end{align}
where we encounter the dimensionless gauge boson, top quark, and Higgs boson
masses as defined in \Eqref{eq:dimlessmuSU2}.

As detailed in the Appendix, ambiguities can arise for the derivation of the
threshold behavior from the choice of the relative orientation between the scalar
vev in fundamental color space and the background color field in adjoint color
space. These ambiguities correspond to slightly different definitions of the
gauge coupling. Furthermore, the computation of the fermionic threshold
behavior contains a minor uncertainty which is not resolved in the present
work. Nevertheless, all qualitative details of the main asymptotic safety
scenario of the present work do not depend on these issues. As a radical
check, we have verified the existence of the fixed point also for a pure
one-loop form of the gauge $\beta$ function.

\subsection{Mass parametrization}
\label{sec:masspar}

Before we turn to an analysis of the flow equations, let us discuss the
redundancy contained in the list of dimensionless mass and coupling parameters
introduced above. For simplicity, we consider the effective potential only to
order $\phi^4$ here; generalizations to higher expansion coefficients
$\lambda_{\geq 3}$ are straightforward. 

A standard viewpoint on the flow equations is the following: as the
matter-field anomalous dimensions are defined by purely algebraic
equations, cf. Eqs.~\eqref{floweq:etaphi}, \eqref{floweq:etaR},
\eqref{floweq:etaLgeneral}, they can be solved as a function of
the dimensionless couplings and the vev $\kappa, \lambda_2, h^2, g^2$
in the SSB regime (analogously in the symmetric regime). Substituting
the anomalous dimensions into the flows of these parameters, yields
the flow equations for the matter sector,
\begin{eqnarray*}
\partial_t \kappa &=& \beta_\kappa(\kappa, \lambda_2, h^2, g^2),\\
\partial_t \lambda_2 &=& \beta_\lambda(\kappa, \lambda_2, h^2, g^2),\\
\partial_t h^2 &=& \beta_h(\kappa, \lambda_2, h^2, g^2), \label{eq:standardbetas}
\end{eqnarray*}
which together with the flow of the gauge coupling \Eqref{floweq:gauge} forms
a consistent and closed set of flow equations. In this standard
parametrization, the dimensionless mass parameters $\mu_{W,\text{t},\phi}$ are
considered as composed out of the couplings and the vev $\kappa$ according to
their definitions \eqref{eq:defmu}. 

It turns out that the alternative viewpoint of formulating the flow in terms
of the mass parameters is particularly useful for the present system in the
SSB regime. For simplicity, let us discuss this viewpoint for SU($\NL=2$),
where the mass parameters are given by $\mu_{W,\text{t},\text{H}}$,
cf. \Eqref{eq:dimlessmuSU2}. In this formulation, we still keep the flow of
the gauge coupling \Eqref{floweq:gauge} understood as a function of these
three mass parameters, cf. \Eqref{floweq:gaugeLSU2}. The flow of the mass
parameters can then be deduced from Eqs.~\eqref{eq:standardbetas} according to
\begin{eqnarray*}
\partial_t \mu_\mathrm{H}^2 &=& 2(\partial_t \kappa) \lambda_2 + 2\kappa
(\partial_t \lambda_2),\\
\partial_t \mu_\mathrm{\text{t}}^2 &=& (\partial_t \kappa) h^2 + \kappa (\partial_t
h^2),\\
\partial_t \mu_\mathrm{W}^2 &=& \frac{1}{2} (\partial_t \kappa) g^2 +
\frac{1}{2} \kappa (\partial_t g^2),
\label{eq:muflowSU2}
\end{eqnarray*}
where $\kappa, h^2, \lambda_2$ in turn are expressed in terms of the mass
parameters on the right-hand side.

\section{Fixed point structure of the model}
\label{sec:FP}

\subsection{Asymptotic safety}

For the construction of an asymptotic safety scenario, see
\cite{Weinberg:1980gg} for reviews, a fixed-point with suitable UV
properties is required. Parametrizing the effective action $\Gamma_k$
by a possibly infinite set of generalized dimensionless couplings
$g_i$, the Wetterich equation provides us with the corresponding
$\beta$ functions $\pat g_i= \beta_{g_i}(g_1,g_2,\dots) $.  A fixed
point $g_i^\ast$ satisfies
\begin{equation}
 \beta_i(g_1^{\ast},g_2^{\ast},...)=0\, ,\ \forall \ i,
\end{equation}
and is called non-Gau\ss ian, if at least one coupling is
nonvanishing $g_j^\ast\neq 0$. The properties of the fixed point can be
quantified by its critical exponents. Linearizing the flow in the fixed-point
regime, 
\begin{equation}
 \partial_t g_i = B_i{}^j  (g_j-g^\ast_j)+\dots, \quad B_i{}^j =\frac{\partial
   \beta_{g_i}}{\partial g_j} \Big|_{g=g^\ast},
\label{eq:lin}
\end{equation}
we obtain the critical exponents $\theta_i$ as the eigenvalues
of the negative Jacobian $(-B_i{}^j)$, which corresponds to the stability matrix. All
eigendirections with $\theta_i<0$ decrease rapidly towards the IR and thus are
irrelevant for the long-range physics.  All relevant directions with exponents
$\theta_i>0$ grow rapidly towards the IR and thus dominate the long range
observables. The marginal directions with $\theta_i=0$ need to be further
classified into marginally-irrelevant and marginally-relevant directions
according to their behavior at higher orders in the expansion about the fixed
point; e.g. the asymptotically free gauge coupling is marginally relevant.  In
total, the number of relevant and marginally-relevant directions equals
the number of physical parameters to be determined by measurements. An
asymptotically safe theory has predictive power, if this number is finite. 
At the Gau\ss ian fixed point $g_i^\ast=0$, this discussion agrees with
the standard perturbative power-counting analysis with the critical
exponents being identical to the canonical dimensions of the couplings. 

A theory allows for an asymptotically safe UV completion, if a non-Gau\ss ian
fixed point with a finite number of UV attractive (relevant or
marginally-relevant) directions exists. Then, renormalized trajectories
emanating from the fixed point can be constructed that correspond to quantum
field theories valid to arbitrarily high energy scales. 

\subsection{Parameter constraints}

As we are working within a truncated theory space, we have to make sure that
possible fixed points do not correspond to mere artifacts of our
truncation. The following discussion of constraints follows that of
\cite{Gies:2009sv}.  The use of a derivative expansion as our expansion scheme
suggests that the quality of the expansion can be deduced from the
quantitative influence of higher derivative operators onto the flow of
leading-order operators. In our truncation, the leading-order effective
potential as well as the Yukawa coupling receive higher-order contributions
only through the anomalous dimensions. Therefore, convergence of the
derivative expansion requires
\begin{equation}
 \eta_{\text{L}}, \eta_{\text{R}}, \eta_\phi \lesssim \mathcal O(1).\label{eq:validity}
\end{equation} 
A similar constraint for the gauge-field anomalous dimension $\eta_W$ is
automatically satisfied as long as we consider only weak gauge couplings. 
Any fixed-point violating \Eqref{eq:validity} is likely to be an artifact of
the present truncation. 

Further constraints arise from the form of the effective potential. In the
symmetric regime, $u$ should exhibit a minimum at vanishing field and should
be bounded from below.  In the polynomial expansion, these criteria translate
into 
\begin{equation*}
m^2,\,  \lambda_{n_{\text{max}}}>0, \label{eq:constraint1}
\end{equation*}
where $n_{\text{max}}$ denotes the highest order taken into account in a
truncated polynomial expansion of the effective potential.  Analogously, the
SSB regime requires a positive minimum, $\kappa>0$, the potential should
again be bounded. In addition, the potential at the minimum must have positive
curvature implying
\begin{equation*}
\kappa, \, \lambda_{n_{\text{max}}},\, \lambda_{2}>0.\label{eq:constraint2}
\end{equation*}
Hermiticity of the Minkowskian action or Osterwalder-Schrader positivity of
the Euclidean action requires
\begin{equation*}
h^2>0, \quad g^2>0,\label{eq:hcrit}
\end{equation*}
which physically is related to Dyson's vacuum stability argument. 

In the mass parametrization for the SSB regime, the above constraints
automatically imply that all mass parameters are non-negative,
$\mu_{W,\text{t},\phi}^2 \geq 0$. From within the mass parametrization, an
independent though somewhat weaker argument can be derived from the general
convexity property of the flow equation \cite{Litim:2006nn}. For the
linear regulator used in this work, convexity implies that
$\mu_{W,\text{t},\phi}^2 > -1$ which physically implies that the regularized
propagators have a finite gap. 

\subsection{Non-Gau\ss ian matter fixed point at finite gauge
  coupling}\label{sec:four}
 
Here and in the following, we exclusively concentrate on
SU($\NL=2$). For illustrative purposes, we will mainly consider
  the simplest case of one fermion generation $\Ng=1$. Though this
  model in isolation would have a global Witten anomaly,
  cf. footnote~\ref{foot:anomaly}, it is phenomenologically more
  reminiscent to the top-Higgs sector of the standard model. In
  App.~\ref{sec:Ng2}, we verify explicitly that the properties of the
  anomaly-free $\Ng=2$ model are essentially identical to the results
  discussed in the following. In order to get an intuition for the
flow equations, we start with the standard formulation in terms of the
couplings and the effective potential. The case of exactly vanishing
gauge coupling was investigated intensively in \cite{Gies:2009sv}. In
fact, the only possible fixed point is the Gau\ss ian gauge fixed
point at $g^2=0$ in the weak gauge-coupling regime.  However, as this
fixed point is only approached asymptotically, let us first search for
possible non-Gau\ss ian fixed point structures in the matter sector,
keeping the gauge coupling fixed at a given value.  For simplicity, we
truncate the effective potential at $\phi^4$ level.

\begin{figure}[!t] 
\begin{center}
 \includegraphics[width=0.22\textwidth]{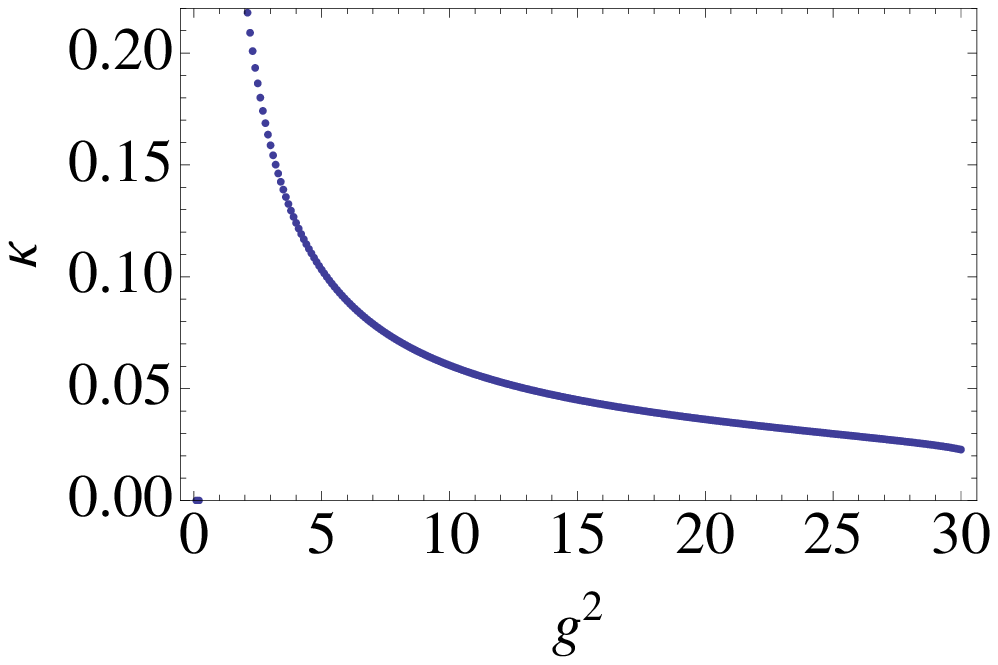}
 \hspace{0.2cm}
 \includegraphics[width=0.22\textwidth]{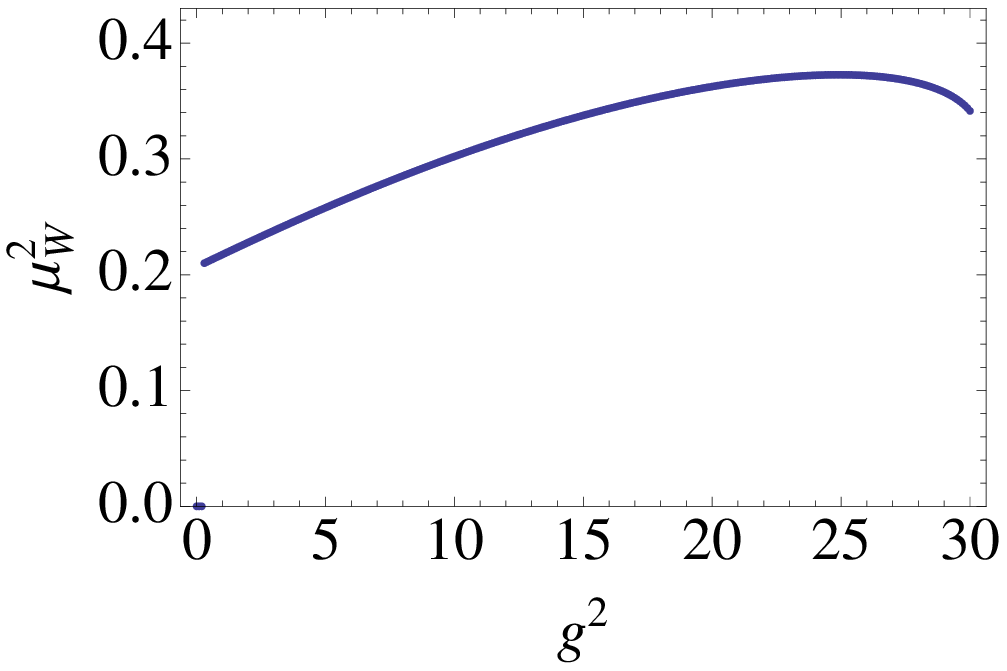}
 \hspace{0.2cm}
 \includegraphics[width=0.22\textwidth]{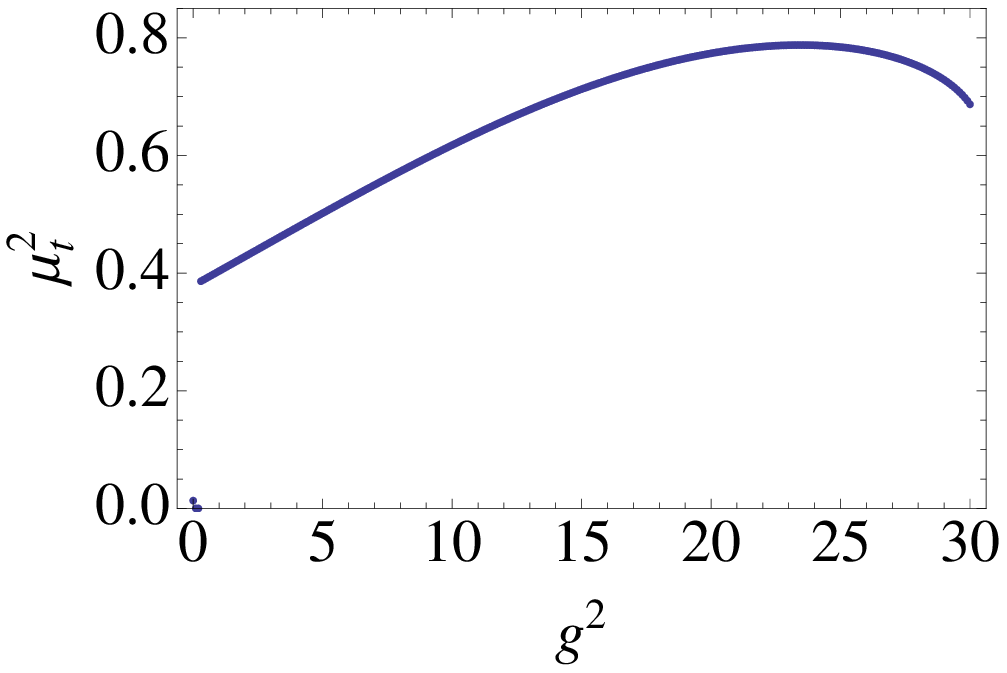}
 \hspace{0.2cm}
 \includegraphics[width=0.22\textwidth]{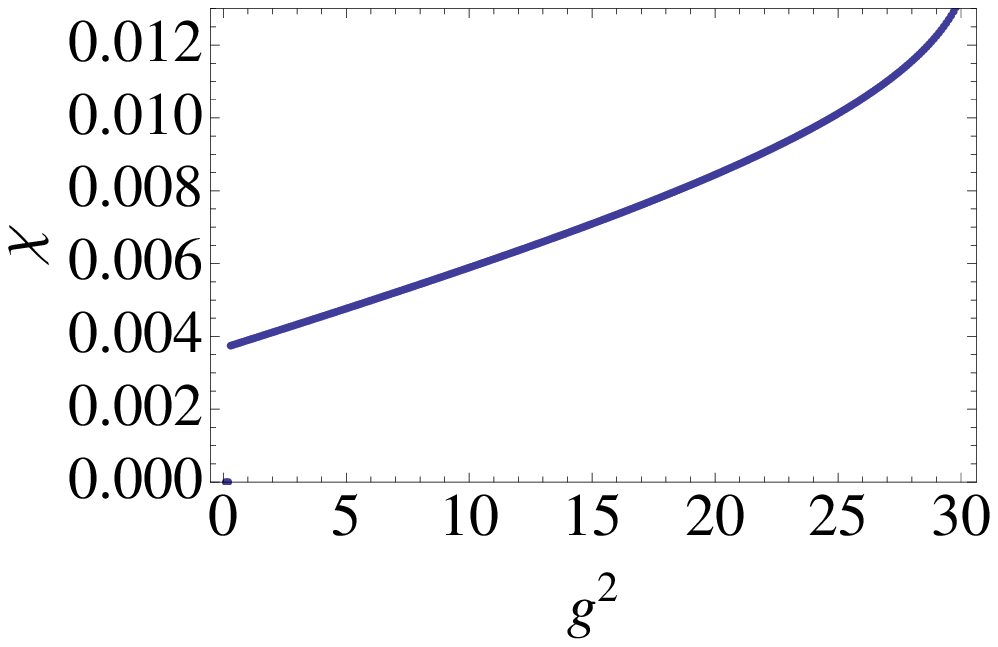}
 \caption{Non-Gau\ss ian matter fixed points for $\NL=2$ as a function of an
   artificially fixed gauge coupling. Whereas the vev $\kappa$ diverges for
   $g^2\to0$, the matter couplings approach zero in such a way that the
   dimensionless mass parameters $\mu_W^2$ and $\mu_{\text{t}}^2$ and the ratio
   $\chi=\mu_{\text{H}}^2/g^2$ tend to finite values, parametrizing a
    true fixed point $\spFP$ of the full system.
}
\label{fig:NLOfixedppoints}
\end{center}
\end{figure}

In fact, for a given finite value of $g^2$, the matter system shows a
non-Gau\ss ian fixed point in the SSB regime. Note that this fixed point is
different from the one found at the leading-order of the derivative expansion in
\cite{Gies:2009sv}. Most importantly, the present fixed point is not
destabilized at next-to-leading order in the derivative expansion. As an
interesting feature, we observe a strong dependence of the fixed point values
on the gauge coupling. In particular, the position of the minimum $\kappa$
diverges with $g^2\to 0$, whereas the Yukawa coupling as well as the scalar
self-interaction approach zero in the same limit. We observe that the
dimensionless gauge and top mass parameters $\mu_W^2, \mu_{\text{t}}^2$
approach finite fixed point values for $g^2\to 0$, whereas the dimensionless
Higgs mass parameter $\mu_{\text{H}}^2$ vanishes. The latter feature implies
that the effective fixed-point potential becomes exceedingly flat for $g^2\to
0$. The ratio $\chi=\mu_{\text{H}}^2/g^2$ approaches a finite constant in this
limit, implying that $\lambda_2$ decreases  $\sim g^4$ at the fixed point. 
These results are depicted in Fig.\ref{fig:NLOfixedppoints}. The fixed
point associated with the limit $g^2\to 0$ will be called $\spFP$ in
the following.

At first sight, this fixed point in the limit $g^2\to 0$ seems to be
identical to the Gau\ss ian fixed point, as all couplings vanish and only
massive free particles remain. This conclusion is, however, wrong for a number
of reasons: first, the true Gau\ss ian fixed point of the present model has
massless gauge bosons and massless chiral fermions and massless or massive
scalar excitations satisfying U($\NL$) symmetry. Second, the dimensionless
mass parameters observed above arise from a subtle interplay of the
interaction terms in the flow equations in the weak-coupling limit; they are a
genuine interaction effect. Third, a massive Gau\ss ian fixed point would
not only violate the symmetries, but also permit any value for the
dimensionful masses which would correspond to fixed scales. In our case, the
dimensionless mass parameters approach fixed points and thus do not define any
scale. Fourth, the critical exponents of this fixed point, computed below, do
not agree with the canonical dimensions at the Gau\ss ian fixed points. 

So far, we have determined the fixed points of the pure matter system for a
given finite gauge coupling. Strictly speaking, the curves shown in
Fig.~\ref{fig:NLOfixedppoints} do note correspond to fixed points except for the
points at $g^2\to 0$, where also the gauge coupling has a fixed point. 
In order to analyze this fixed point more properly, we now turn to the mass
parametrization introduced in Sect.~\ref{sec:masspar}.

\subsection{Non-Gau\ss ian fixed points in the mass parametrization}

Let us analyze the fixed-point structure on the basis of the matter flow
equations in mass parametrization Eqs.~\eqref{eq:muflowSU2} read together with
the gauge coupling flow. From the preceding analysis, we already infer that
the ratio
\begin{equation}
\chi=\frac{\mu_{\text{H}}^2}{g^2}, \label{eq:defchi}
\end{equation}
approaches a constant at the desired fixed point. Hence, we consider
the right-hand sides of these flow equations in the limit $g^2\to 0$
for finite $\mu_{\text{t}}^2, \mu_W^2, \chi$. In this limit, the fixed
point conditions
\begin{eqnarray}
\partial_t g^2 &=& 0,\quad
\partial_t \mu_\mathrm{H} = 0,\quad
\end{eqnarray}
are automatically satisfied. Nontrivial information remains encoded in the
flow of the new variable $\chi$,
\begin{equation}
\partial_t \chi^2 = \frac{1}{g^2} \pat \mu_{\text{H}}^2 -
\frac{\mu_{\text{H}}^2}{g^4} \pat g^2, \label{eq:chiflow}
\end{equation}
the $g^2\to 0$ limit of which remains finite on the right-hand side except for 
the fixed points computed below.

As a particular property, we observe that the
flows of $\mu_{\text{t}}^2, \mu_W^2$ become degenerate,
\begin{equation}
\partial_t \mu_{\mathrm{t}}^2 = \frac{\mu_{\mathrm{t}}^2}{\mu_W^2}(\partial_t
\mu_W^2). \label{eq:mutflow}
\end{equation}
This degeneracy has an important consequence: possible fixed points
for the remaining matter sector parametrized in terms of the three
variables $(\chi,\mu_{\text{t}}^2, \mu_W^2)$ have to be determined
from only two independent equations, Eqs.~\eqref{eq:chiflow}, and
\eqref{eq:mutflow}. This implies that for any non-trivial fixed point
a degenerate one-parameter family, i.e., a line of fixed points has to
exist. Since the degeneracy encoded in \Eqref{eq:mutflow} becomes
exact only in the limit $g^2\to0$ (together with $\mu_{\text{H}}^2 \to
0$ and $\mu_W^2$, $\mu_{\text{t}}^2$, $\chi$ finite), this line of
fixed points remains invisible in the standard parametrization by
approaching the fixed point $g_\ast^2=0$ from finite values of $g^2$.

\begin{figure}[ht!]
\begin{center}
 \psfrag{x}{\tiny{$\mu_\mathrm F^\ast$}}
 \psfrag{y}{\tiny{$\mu_\mathrm G^\ast$}}
 \includegraphics[width=0.23\textwidth]{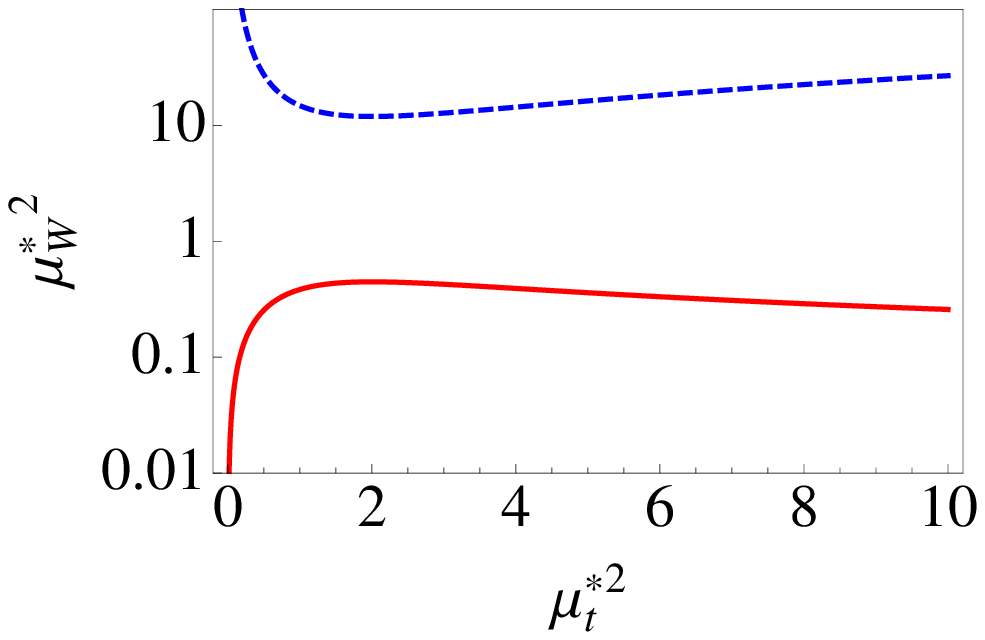}
 \hspace{0.2cm}
 \psfrag{x}{\tiny{$\mu_\mathrm F^\ast$}}
 \psfrag{y}{\tiny{$\chi^\ast$}}
 \includegraphics[width=0.23\textwidth]{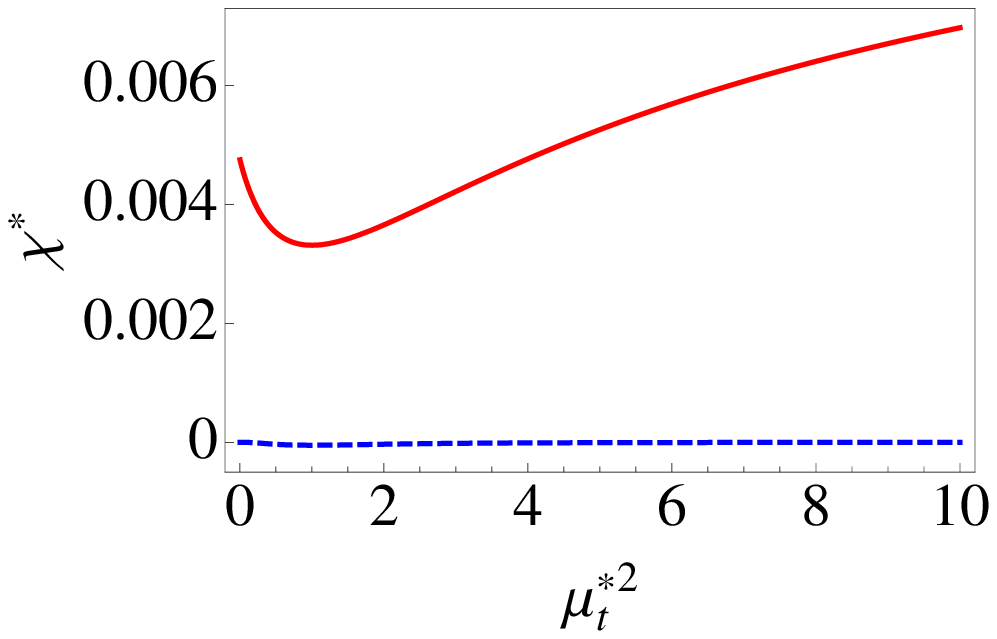}
 \caption{Fixed point values for $\mu_W^2$ (left panel) and $\chi$ (right
   panel) as a function of the fixed point value of $\mu_{\mathrm t}^2$ for
   $\NL=2$. Our main solution is shown as a solid line, whereas the dashed
   line depicts a second solution with legitimate values of the gauge and
   fermion sector, but a (physically inadmissible) negative $\chi$.}
\label{fig:fixedpointsNL2}
\end{center}
\end{figure}

Knowing already from the preceding standard analysis that a fixed
point exists, let us now search for the corresponding line of fixed
points within the mass parametrization, by looking for fixed point
solutions satisfying $\pat\chi=0$ and $\pat \mu_{\text{t}}^2=0$. 
The reduced flow equation for the top mass parameter in the above mentioned
limit reads
\begin{equation}
\partial_t \mu_{\mathrm{t}}^2 =  -2\mu_{\mathrm{t}}^2  
+ \frac{\mu_{\mathrm t}^2}{8\pi^2 \chi}\left( \frac{9}{4(1+\mu_W^2)^2} 
- \frac{\mu_{\mathrm t}^2}{\mu_W^2 (1+\mu_{\mathrm t}^2)^2} \right).
\label{eq:redmutflow}
\end{equation}
The correspondingly reduced fixed point equation 
 $\pat \chi=0$ yields exactly one solution, reading
\begin{equation}
\chi^\ast= -\frac{1}{16\pi^2}\Big(\frac{ \mu_{\mathrm{t}}^2 (1+3 \mu_{\mathrm{t}}^2)}{(1+\mu_{\mathrm{t}}^2)^3 \mu_W^2}-\frac{9(1+3\mu_W^2)}{4(1+\mu_W^2)^3}\Big).
\end{equation}
We can plug this result into the flow equation \eqref{eq:redmutflow} for
$\mu_{\mathrm{t}}^2$ to solve for fixed points, giving three solutions for
$\mu_W^2$ as a function of $\mu_{\mathrm{t}}^2$. The solutions with positive,
real masses $\mu_{\mathrm t}^2$ and $\mu_W^2$ are depicted in
Fig.\ref{fig:fixedpointsNL2}. Apart from our main solution (solid line), the
second solution (dashed lines) results in negative values for $\chi^\ast$ and
hence also for $\mu_{\mathrm H}^2$, and therefore will be discarded in the
remainder.  As the flows of $\mu_\mathrm{t}^2$ and $\mu_W$ are proportional, 
their ratio remains undetermined and thus parametrizes a line of fixed points
as expected above.

It is reassuring to observe that the fixed point $\spFP$ identified in the standard
parametrization of the flow equations in Subsect.~\ref{sec:four} in the limit
$g^2\to 0$,
\begin{equation}
\spFP: \quad (\mu_{\mathrm t}^{\ast 2}, \mu_W^{\ast 2}, \chi^\ast) \simeq (0.38,
0.21,0.0037),
\label{eq:specialFP}
\end{equation}
cf. Fig.\ref{fig:NLOfixedppoints}, is exactly on the (physically
admissible) line of fixed points. Along this line, we observe
  that for $\mu_{\mathrm t}^{\ast 2}=\mathcal{O}(1)$, we have
  $\mu_{W}^{\ast 2}\simeq\mathcal{O}(0.1, \dots ,1)$, but a much smaller
  $\chi^\ast =\mathcal{O}(10^{-3})$. These properties also affect the
  typical mass hierarchy in the IR, see below.

\begin{figure}[h!]
\begin{center}
 \psfrag{x}{\tiny{$\mu_\mathrm F^\ast$}}
 \psfrag{y}{\tiny{Re[$\theta_i$]}}
 \includegraphics[width=0.23\textwidth]{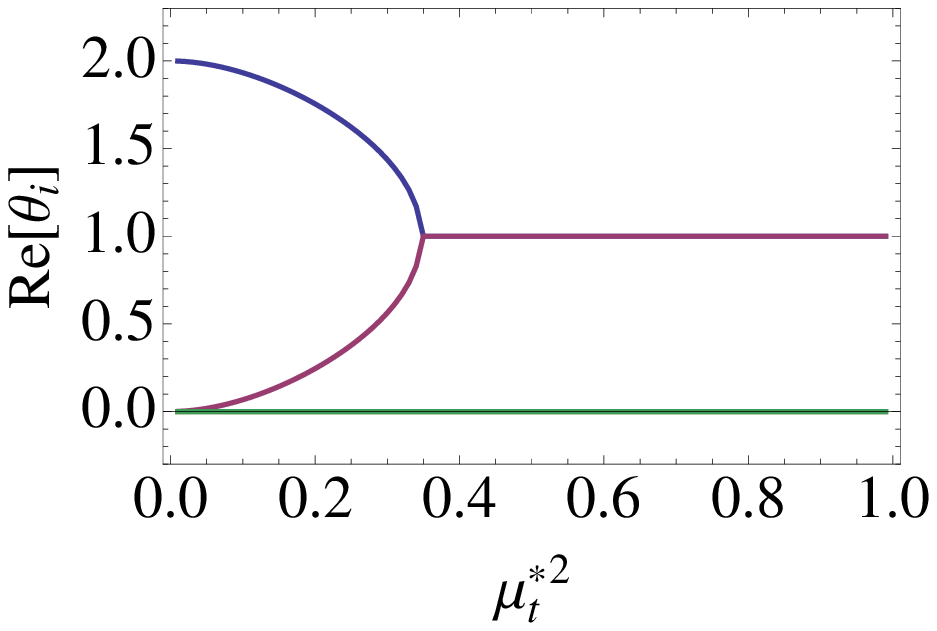}
 \psfrag{x}{\tiny{$\mu_\mathrm F^\ast$}}
 \psfrag{y}{\tiny{Im[$\theta_i$]}}
 \includegraphics[width=0.23\textwidth]{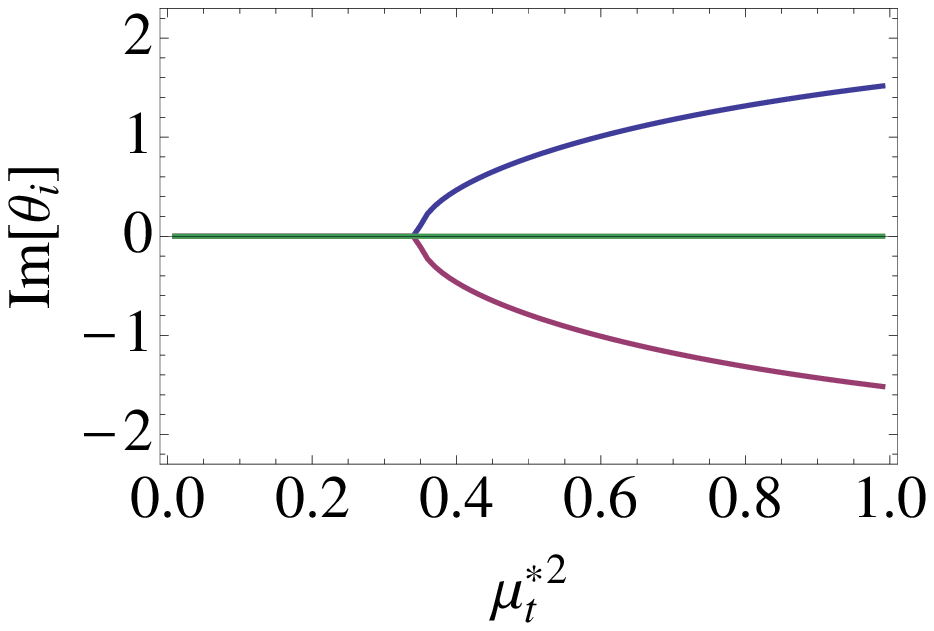}
 \caption{Critical exponents for the line of fixed points computed in
   the mass parametrization as a function of the fixed point top mass
   parameter $\mu_{\text{t}}^{\ast 2}$ for $\NL=2$; left panel: real
   parts, right panel: imaginary parts. The fixed point near
     $\mu_{\mathrm t}^{\ast 2}\simeq 0.35$, where the two positive
     exponents merge and start to form a complex pair, corresponds to
     the fixed point $\mathcal{B}$ introduced below.}
\label{fig:thetaNL2}
\end{center}
\end{figure}

Let us finally compute the critical exponents $\theta_i$ along the line of
fixed points from the stability matrix, as defined in \Eqref{eq:lin}, in terms
of the generalized couplings $ g_j$ within the mass parametrization
$\{g^2,\mu_{\mathrm{t}}^2,\mu_W^2,\chi\}$. The results for the critical
exponents are depicted in Fig. \ref{fig:thetaNL2} as a function of the fixed
point value $\mu_{\text{t}}^{\ast 2}$. Two of the four critical exponents are
zero. Two others start off at $\theta_{1}=2$, $\theta_{2}=0$, respectively,
for $\mu_{\text{t}}^{\ast 2}=0$, i.e., at the same values as the canonical
dimensions at the Gau\ss ian fixed point, but then approach each other for
finite $\mu_{\text{t}}^{\ast 2}$, merging at $\mu_{\text{t}}^{\ast 2}\simeq
0.35$. For
$\mu_{\text{t}}^{\ast 2}  > 0.35$ they form a complex pair with equal real
part, $\text{Re}\theta_{1,2}=1$ and conjugate imaginary parts, as shown in the
right panel of Fig. \ref{fig:thetaNL2}.

As a particular example, let us list the critical exponents at the
fixed point $\spFP$ already discovered in the standard
parametrization of the flow given in \Eqref{eq:specialFP}. These
exponents are given by
\begin{equation}
\spFP: \quad \theta_{1/2} = 1 \pm 0.36  i \, , \theta_3 = \theta_4 = 0 \, .
\end{equation}
The first two critical exponents form a complex pair and thus describe a
spiraling approach towards the fixed point on the corresponding UV critical
hypersurface. One of the marginal directions points along the direction of the
gauge coupling. From asymptotic freedom of the gauge sector, we can infer that
this is a marginally-relevant direction. The other marginal direction is
related to the existence of a line of fixed points. This direction must be
exactly marginal in our truncation, as a perturbation of the couplings at a
given fixed point along the line of fixed points just puts the system onto
another fixed point where the flow vanishes completely. 

Another special example is the fixed point $\mathcal{B}:\,\,
  (\mu_{\mathrm t}^{\ast 2}, \mu_W^{\ast 2}, \chi^\ast) \simeq (0.35,
  0.19,0.0037)$, denoting the
  branch point where the two largest real exponents start to form a
  complex pair. Exactly at $\mathcal{B}$, we have $\quad \theta_{1/2}
  = 1$, $\theta_{3,4} = 0$.

To summarize, we have identified a line of UV stable fixed points that can
serve to define UV complete quantum field theories of gauged Higgs-Yukawa
models by means of suitable RG trajectories that emanate from the fixed point
in the UV. Specifying a certain trajectory yields a fully predictive
long-range theory. It is instructive to compare our asymptotically safe models
with the standard Gau\ss ian fixed point used for perturbative
renormalization; theories near the Gau\ss ian fixed points have to
be defined in terms of four physical parameters corresponding, for instance,
to the relevant mass parameter of the Higgs potential, and the marginal scalar
self-interaction, Yukawa coupling and gauge coupling. 

By contrast, in the present asymptotically safe model defined in terms
of a trajectory emanating from a fixed point, each trajectory is
defined in terms of three physical parameters corresponding to the two
relevant directions and the marginally relevant gauge coupling. The
fourth parameter does not correspond to a physical parameter in a
particular theory, but corresponds to choosing one fixed point on the
line of fixed points, i.e., choosing among a one-parameter family of
theories. Of course, this difference in the conceptual meaning of
parameters is not substantial from a pragmatic viewpoint, as in total
four different measurements are needed in order to fix the long-range
behavior of the system unambiguously. 

Whether or not the existence of a line of fixed points also holds
beyond our truncation cannot be told from the present
investigation. It is at least well conceivable that the degeneracy
manifested by \Eqref{eq:mutflow} is lifted by the influence of higher
order operators. If so, only a finite number of fixed points might
remain, thus potentially also reducing the number of physical
parameters as observed in the ungauged Yukawa models
\cite{Gies:2009hq, Gies:2009sv}.

\section{Flow from the ultraviolet to the electroweak scale}\label{sec:six}

For the asymptotically safe models discovered above to be viable
building blocks of the standard model Higgs sector, we need to show
that the UV critical hypersurface contains trajectories that end up in
the Higgs phase of the model with massive gauge bosons, a massive top
quark and a massive scalar Higgs boson. This is indeed the case as we
demonstrate in the following by integrating the RG flow for different
sets of initial conditions towards the long-range physics.

Since the fixed points exhibit positive critical exponents of order
$\mathcal{O}(1)$, we face the standard technical problem that initial
conditions near the fixed points have to be fine-tuned in order to separate the
UV scale from the scale of IR observables. The necessary fine-tuning in the
Yukawa sector is complicated by the marginally-relevant gauge sector which has
to be UV adjusted as to provide for a suitable IR coupling strength. This
suitable UV adjustment is not a conceptual problem, but merely a tedious
search for the desired initial conditions which can straightforwardly be
solved, e.g., by suitable bisection techniques. In the present case, this
problem is slightly complicated by the fact that the line of fixed points only
becomes visible in the limit $g^ 2\to0$. In order to ensure that the flow is
started sufficiently near the line of fixed points, we thus have to start in
the deep UV where $g^2$ is sufficiently small.

\begin{figure}[t!]
\begin{center}
\includegraphics[width=0.23\textwidth]{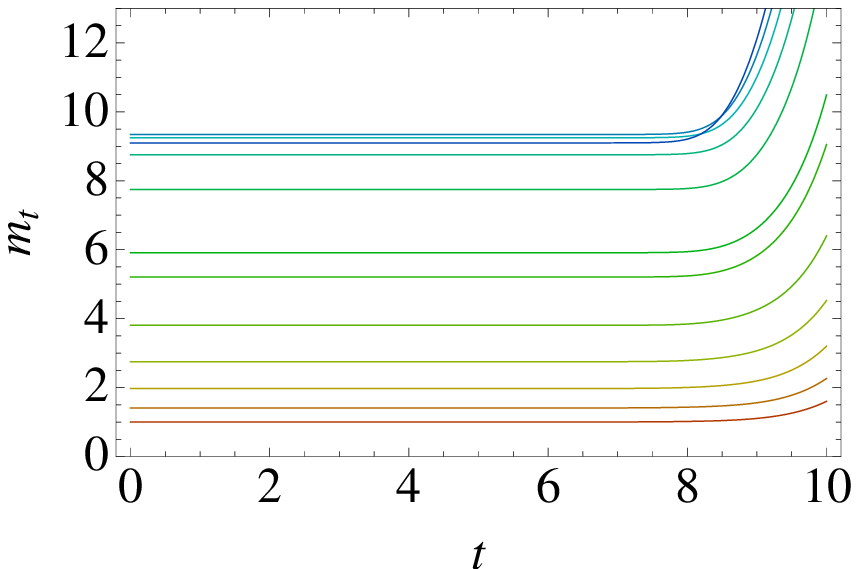}
\includegraphics[width=0.23\textwidth]{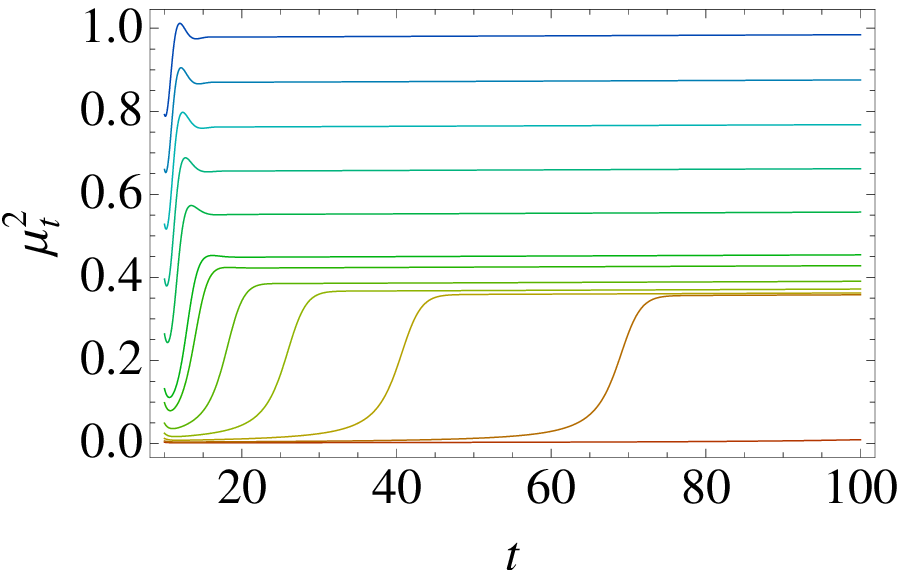}
\caption{Right plot: typical UV flows to the line of fixed points for
  the dimensionless parameter $\mu_t^2$. Left plot: corresponding IR
  trajectories for the dimensionful top mass with freeze-out. The set
  of curves corresponds to different values of $\mu_t^2$ at the
  intermediate scale $t=10$ (larger to smaller from top to bottom)
 and are normalized such that the smallest initial $\mu_t^2$
    value yields a top mass $m_{\mathrm{t}}=1$ in arbitrary units. }
\label{fig:flowstypUVandIR}
\end{center}
\end{figure}

For simplicity, we will choose such a small value in the following
flow examples. 
We set $g^2=1/2000$ at an arbitrary initialization scale $t_{\text{init}}=\ln k/\Lambda=10$. 
In the present work, this scale has a meaning of an intermediate scale, separating the infrared $t<10$, where the IR observables and the mass spectrum are built up, from the flow towards the fixed point regime for $t>10$. 
Typical flows ending up in the Higgs phase are shown in Fig.~\ref{fig:flowstypUVandIR}. 
All these flows are initiated near the line of UV fixed points starting
with different values of $\mu_{\text{t}}^2$ within the interval $[10^{-5},0.8]$ at $t_{\text{init}}=10$. 
For the remaining two variables, $\mu_W^2$ and $\chi$, we employ initial values that lie close to the fixed point corresponding to the current choice of $\mu_{\text{t}}^2$, cf. Fig.~\ref{fig:fixedpointsNL2}. 
In the deep UV (right panel), the flows approach the line of fixed points at
different values of $\mu_{\text{t}}^{\ast 2}$ (and corresponding
values of $\mu_W^{\ast 2}$ and $\chi^\ast$). Towards the IR, the flows
rapidly freeze out, implying that the dimensionful mass parameters
such as $m_{\text{t}}$ approach finite values (left panel). The units are arbitrary and chosen such that the lowest trajectory yields an IR value of 1 for the top mass $m_{\text{t}}$.  We
observe a very similar behavior for the other two mass variables, with
the dimensionless parameters $\mu_W^2$ and $\chi$ approaching their
corresponding fixed points in the UV and the dimensionful counterparts
$m_{\text{W}}$ and $m_{\text{H}}$ freezing out at finite values in the
IR. The gauge coupling exhibits the perturbative asymptotically free
log-running including the slight threshold modifications in the UV, whereas
the log-running is more strongly modified in the IR because of the
decoupling of massive modes. 

These flows illustrate our conclusion that the UV line of fixed points render the
present gauged Higgs-Yukawa model asymptotically safe. Whereas the IR exhibits
a standard Higgs phase indistinguishable from the perturbative standard
scenario, the UV is controlled by a fixed point at which the continuum limit
can be taken. The theory thus is UV complete. 

In addition to these results which can already be anticipated from the
pure fixed point analysis, we observe further aspects of our model
which require the solutions of the flow equation: for flows starting
at generic values of $\mu_{\text{t}}^2,\mu_W^2, \chi$ near the line of
fixed points, the hierarchy of the fixed point values is transmuted
into a similar hierarchy of the particle masses: e.g., for typical
flows that approach a fixed point with $\mu_{\text{t}}^2$ of order
$\mathcal{O}(0.1, \dots ,1)$ 
, we read off from
Fig.~\ref{fig:fixedpointsNL2} (right panel), that the corresponding
$\chi$ values are typically two orders of magnitude smaller, whereas
$\mu_W^2$ is also of order $\mu_{\text{t}}^2$ in this regime (left
panel). The same hierarchy then generically remains present in the
mass spectrum: the top and $W$ mass are of the same order, whereas the
Higgs mass is typically about two orders of magnitude smaller.

\begin{figure}[t!]
\begin{center}
\includegraphics[width=0.23\textwidth]{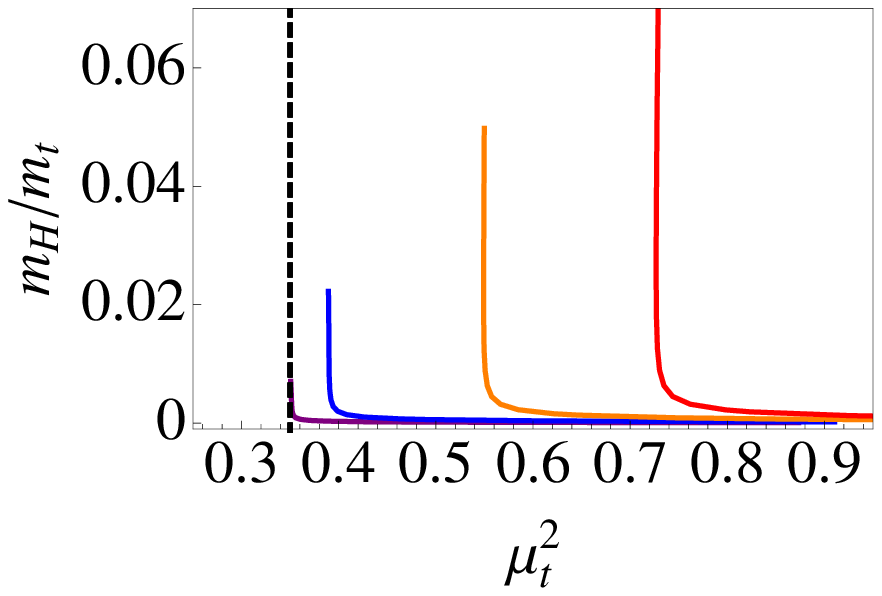}
\includegraphics[width=0.23\textwidth]{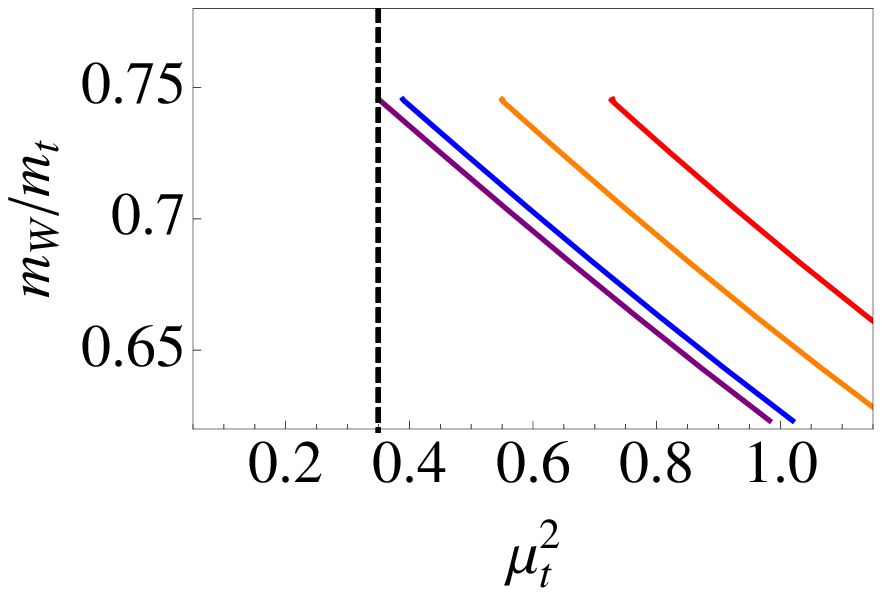}
   \caption{Compilation of IR results for flows starting on the line
     of fixed points, as a function of the dimensionless mass
     parameter $\mu_{\text{t}}^2$ in the deep UV for four different
     starting values of the gauge coupling
     $g^2\in\{10^{-4},5\cdot10^{-5},10^{-5},10^{-6}\}$ (red, orange,
     blue and purple, respectively) at $t=10$. The deep UV value of
     $\mu_{\text{t}}^2$ is read off at a rather extreme
     reference scale $t=30000$. Whereas the $W$ to top mass ratio
     remains within a very narrow range by choosing natural initial
     conditions (right panel), realistic Higgs-to-top mass ratios can
     be approached for specifically tuned flows (left panel). In
       the limit of weak gauge coupling, these tuned flows appear to
       be connected with the branching fixed point $\mathcal{B}$
       (dashed lines).}
\label{fig:massratios}
\end{center}
\end{figure}

This preservation of UV-IR hierarchy is lifted, if the system is tuned
to certain values which depend a bit on the value of the gauge
coupling at the initialization scale $t_{\text{init}}$. For these
specifically tuned flows, the UV behavior still exhibits the parameter
hierarchy, whereas the IR can behave differently. A variety of UV-IR
solutions for varying initial condition $\mu_{\text{t}}^2$ at
$t_{\text{init}}$ are compiled in Fig.~\ref{fig:massratios} and
depicted as a function of the deep UV value for $\mu_{\text{t}}^2$
(read off at $t>130$, cf. Fig. \ref{fig:flowstypUVandIR}). These
initial conditions have been chosen such that the $W$ mass to top mass
ratio remains slightly above the physical value near $\simeq
  0.5$, as shown in the right panel. This implies that, as the top
mass increases by a factor of two, also the $W$ mass is about twice as
large. This is different for the Higgs mass (left panel), which
generically is two orders of magnitude smaller than the top mass. Only
for the particularly tuned trajectories, the situation changes and we
find Higgs mass to top mass ratios ultimately approaching the
realistic ratio $125/175\simeq 0.7$. Four such trajectory sets with
corresponding tuning are shown in Fig.~\ref{fig:massratios} for
different values of the gauge coupling
$g^2\in\{10^{-4},5\cdot10^{-5},10^{-5},10^{-6}\}$ at
$t_{\text{init}}$.

The particularities of these specifically tuned flows become obvious
from the behavior of the trajectories at intermediate scales. Whereas
generic flows consist of a UV fixed-point regime, an IR freeze-out
regime and a cross-over in between, flows that lead to larger Higgs
masses exhibit an additional quasi-fixed-point regime at intermediate
scales. This {``walking'' behavior} is, for instance, visible for
the lower lines in Fig.~\ref{fig:flowstypUVandIR} (right panel),
showing a cross-over from the UV fixed point to an intermediate {walking}
regime which can extend over a wide range of scales. This intermediate
regime is characterized by small values of the dimensionless mass
parameters. The corresponding $\beta$ functions are obviously small
here but non-vanishing. We interpret this {walking} regime as the remnant of the
line of the ($g^2\to 0$) fixed points at finite values of $g^2$.

It is interesting to observe that trajectories exhibiting this
{walking} regime are connected with sharply defined fixed
point values of $\mu_{\text{t}}^{\ast 2}$. The latter in turn depends
on the initial choice for the value of the gauge coupling. In the
limit of small initial gauge coupling, our results indicate that the
trajectories approach the branching fixed point $\mathcal{B}$ with
$\mu_{\text{t}}^{\ast 2} \simeq 0.35$, cf. dashed lines in
Fig.~\ref{fig:massratios}. In our parameter scans, we have not been
able to find trajectories that emanate from the line of fixed points
at values $\mu_{\text{t}}^{\ast 2}<0.35$. This might indicate that the
RG flow close to this branch of the line of fixed points is rather
strong. This would go along with the fact that the value of the
largest critical exponent increases for smaller $\mu_{\text{t}}^{\ast
  2}$, cf. Fig.~\ref{fig:thetaNL2} (left panel), such that finding
those trajectories requires more sophisticated fine-tuning
procedures. Alternatively, these fixed points might not be directly
connected to a massive Higgs phase in the IR. These questions deserve
further study.

\section{Conclusions}
\label{sec:conc}

We have discovered a line of interacting fixed points in the RG flow
of gauged chiral Higgs-Yukawa models. Each fixed point gives rise to a
novel universality class of UV complete asymptotically safe quantum
field theories with interacting fermions, gauge fields and elementary
scalars. We have demonstrated that UV complete RG trajectories
emanating from the line of fixed points exhibit a Higgs phase with
massive top quark, gauge bosons and Higgs boson. If similar properties
hold for the standard model, the existence of such a line of fixed
points solves the triviality problem of the top-Higgs sector of the
standard model.

The non-trivial UV behavior is characterized by asymptotic freedom in
all interaction couplings and a quasi-conformal behavior in all
mass-like parameters. In other words, the fixed point theories live in 
the symmetry-broken regime with all masses running proportional to the
RG scale. In particular, the scalar effective potential approaches
asymptotic flatness in the UV, with a non-vanishing minimum increasing
inversely proportional to the asymptotically free gauge coupling.

Our computations are based on a functional RG approach at
next-to-leading order in a derivative expansion. Our results for the
UV behavior at leading order and next-to-leading order are identical,
providing evidence for a good convergence of our nonperturbative
approximation scheme. In this scheme, we have determined the critical
exponents along the line of fixed points, two of which belong to RG
relevant directions. The gauge coupling remains a marginally-relevant
direction, whereas perturbations along the line of fixed points are
exactly marginal. As a consequence, these asymptotically safe theories
have one parameter less than their corresponding (non UV-complete)
perturbative counterparts near the Gau\ss ian fixed point. However,
this one parameter is transmuted into the one-parameter family of fixed
points distinguishing different universality classes along the line of
fixed points.

We have performed non-exhaustive scans of typical long-range
properties of the system in the Higgs phase. A generic feature appears
to be that the mass of the Higgs boson is about two orders of
magnitude smaller than the top quark mass and the gauge boson mass;
the latter masses are typically of similar size. This mass hierarchy
is a consequence of a corresponding hierarchy of fixed point
values. We observe that this hierarchy can be lifted along special
trajectories which exhibit an intermediate walking regime. If similar
properties are also found for the full standard model, a construction
of a realistic model with improved UV behavior seems within reach.

In this respect, the IR results of the present model are
  remarkable, since the ``IR window'', i.e., the physical parameter
  space accessible in the IR, appears to be rather different from that
  of the standard model in a perturbative treatment. The latter as
  well as many of its extensions typically feature upper and lower
  bounds on the mass of the Higgs boson as a function of the UV cutoff
  \cite{Maiani:1977cg} with the recently measured Higgs boson mass
  \cite{Aad:2012tfa} being near or even somewhat below the lower
  bound. The present model therefore provides for an example that a
  modified nonperturbative UV running of the couplings can strongly
  influence the shape of the IR window without modifying the particle 
  content or the interactions.

\section*{Acknowledgments}

We thank Lukas Janssen, Axel Maas, Jan Pawlowski, Ren\'{e} Sondenheimer, Gian Paolo Vacca for
interesting and enlightening discussions. HG, MMS, and LZ acknowledge
support by the DFG under grants GRK1523, Gi 328/5-2 (Heisenberg
program) and FOR723. MMS is supported by the grant ERC- AdG-290623. The work of SR is supported by the DFG within the Emmy-Noether program (Grant SA/1975 1-1).

%
%
%
%

\appendix


\section{Regulators and threshold functions}

\subsection{Regulators} 
We have to evaluate the r.h.s of eq.~\eqref{flowequation}, for which
we need the $\Gamma_k^{(2)}$ matrix.  Let us consider the fields
$\phi_i$, $\psi_\mathrm L$, $\psi_\mathrm R$, $W$, $c$, $\bar{c}$ as
column vectors, with a number of components respectively given by
$\NL$, $d_\gamma \Ng \NL$, $d_\gamma \Ng$, $d \dG$, $\dG$,
$\dG$.  Accordingly let us consider $\bar{\psi}_\mathrm L$ and
$\bar{\psi}_\mathrm R$ as row vectors.  Taking care of the partly
Grassmann-valued field components and of the Fourier conventions, let us
denote by $\Phi^{\mathrm T}(q)$ the row vector with components
$\phi^{\mathrm T}_1(q)$, $\phi^{\mathrm T}_2(q)$, $\psi^{\mathrm
  T}_\mathrm L(q)$, $\bar{\psi}_\mathrm L(-q)$, $\psi^{\mathrm
  T}_\mathrm R(q)$, $\bar{\psi}_\mathrm R(-q)$, $W^{\mathrm T}(q)$,
$c^{\mathrm T}(q)$, $\bar{c}^{\mathrm T}(q)$, and by $\Phi(p)$ the
column vector given by its transposition.  Then $\Gamma_k^{(2)}$ is
computed as follows
\begin{equation*}
 \Gamma_k^{(2)}=
\frac{\overrightarrow{\delta}}{\delta \Phi^{\mathrm{T}}(-p)}
\Gamma_k
\frac{\overleftarrow{\delta}}{\delta \Phi(q)}\,.
\end{equation*}
For a proper IR regularization, a regulator which is diagonal in field space is sufficient and convenient,
\begin{equation*}
R_k(q,p)=\delta(p-q)
\begin{pmatrix}
R_{\mathrm{B}} & 0 & 0 & 0 & 0\\
0 & R_{\mathrm{L}} & 0 & 0 & 0\\
0 & 0 & R_{\mathrm{R}} & 0 & 0\\
0 & 0 & 0 & R_{\mathrm{G}} & 0\\
0 & 0 & 0 & 0 & R_{\mathrm{gh}}\\
\end{pmatrix}\!(p)\ ,
\end{equation*}
with a $2N_\mathrm{L}\times 2N_\mathrm{L}$ matrix for the scalar bosonic sector
\begin{equation*}
R_{\mathrm{B}}(p)=
\begin{pmatrix}
\delta^{ab} & 0\\
0 & \delta^{ab}\\
\end{pmatrix}Z_{\phi}p^2r_{\mathrm{B}}(p^2),
\end{equation*}
an $2d_\gamma \Ng N_\mathrm{L}\times 2d_\gamma \Ng N_\mathrm{L}$ matrix for the left-handed spinor
\begin{equation*}
R_{\mathrm{L}}(p)=-
\begin{pmatrix}
0 & \delta^{ab}\pslash^{\mathrm{T}}\\
\delta^{ab}\pslash & 0\\
\end{pmatrix}Z_{\mathrm{L}}r_{\mathrm{L}}(p^2)\ ,
\end{equation*}
an $2d_\gamma \Ng\times 2d_\gamma \Ng$ matrix for the right-handed spinor
\begin{equation*}
R_{\mathrm{R}}(p)=-
\begin{pmatrix}
0 & \pslash^{\mathrm{T}}\\
\pslash & 0\\
\end{pmatrix}Z_{\mathrm{R}}r_{\mathrm{R}}(p^2)\ ,
\end{equation*}
a $d\dG \times d\dG$ matrix for the gauge vector boson
\begin{eqnarray*}
R_{\mathrm{G}}(p)&=&Z_Wp^2 r_{\mathrm{GT}}(p^2) \Pi_{\mathrm{T}}^{\mu\nu}(p)\delta^{ij}\\
&&+\frac{Z_\phi p^2 r_{\mathrm{GL}}(p^2)}{\alpha}\Pi_{\mathrm{L}}^{\mu\nu}(p)\delta^{ij}\,,
\end{eqnarray*}
where the $\Pi$'s are the usual longitudinal and transverse projectors with respect to $p_\mu$, 
and a $2\dG\times 2\dG$ matrix for the ghosts
\begin{equation*}
R_{\mathrm{gh}}=
\begin{pmatrix}
0 & \delta^{ij}\\
-\delta^{ij} & 0\\
\end{pmatrix}p^2r_{\mathrm{gh}}(p^2)\,.
\end{equation*}
Notice that here and in the whole paper we set $Z_{\mathrm{gh}}=1$ at any scale, that is: we neglect
$\eta_{\mathrm{gh}}$.
Choosing different regulators for the scalar bosons
(B), for the transverse gauge bosons (GT), for the longitudinal gauge boson (GL),
for the ghosts (gh) and for the left-handed (L) as well as for the right-handed (R) spinors,
allows one to write the flow equation in the form
\begin{equation*}
\partial_t \Gamma_k=\frac{1}{2}\tilde \partial_t{\rm STr}\log(\Gamma_{k}^{(2)}+R_k)\,,
\end{equation*}
where
%
%
%
\begin{eqnarray}\label{eq:tildedt}
\tilde\partial_t&\equiv& \frac{\partial_t(Z_\phi r_\mathrm B)}{Z_\phi}\!\cdot\!\frac{\delta}{\delta r_\mathrm B}
+\frac{\partial_t(Z_\mathrm L r_\mathrm L)}{Z_\mathrm L}\!\cdot\!\frac{\delta}{\delta r_\mathrm L}\\ \nonumber
&&+\frac{\partial_t(Z_\mathrm R r_\mathrm R)}{Z_\mathrm R}\!\cdot\!\frac{\delta}{\delta r_\mathrm R}
+\frac{\partial_t(Z_W r_{\mathrm{GT}})}{Z_W}\!\cdot\!\frac{\delta}{\delta r_{\mathrm{GT}}}\\ \nonumber
&&+\frac{\partial_t(Z_\phi r_{\mathrm{GL}})}{Z_\phi}\!\cdot\!\frac{\delta}{\delta r_{\mathrm{GL}}}
+\partial_t r_{\mathrm{gh}}\!\cdot\!\frac{\delta}{\delta r_{\mathrm{gh}}}
\end{eqnarray}
%
%
%
and $\cdot$ denotes multiplication as well as integration over the common argument of the shape functions of the two factors.
After having performed this differentiation we are free to specify the form of the shape functions $r$.
See App.~\ref{section:threshold} for an example of such a choice.

\subsection{Threshold functions} \label{section:threshold}
 
Since in the SSB regime one of the left-handed Weyl fermions together with the right-handed one gets massive, 
it is useful to introduce a superscript (F) to denote the corresponding Dirac fermion.
Then the regularized kinetic (or squared kinetic) terms are given by
\begin{eqnarray*} 
P_{\mathrm{B}/\mathrm{GT}/\mathrm{GL}/\mathrm{gh}} (x)\!&=&\! 
x(1+r_{\mathrm{B}/\mathrm{GT}/\mathrm{GL}/\mathrm{gh}}(x))\\
P_\mathrm L(x)\!&=&\! x(1 + r_{\mathrm L}(x))^2\\
P_\mathrm F(x)\!&=&\! x(1 + r_{\mathrm L}(x))(1 + r_{\mathrm R}(x)) \,.
\end{eqnarray*}
Accordingly, the loop momentum integrals appearing on the r.h.s. of
the flow equation are classified, implicitly defining the
corresponding threshold functions. 
In the following, the operator $\tilde\partial_t$
is the one defined in \Eqref{eq:tildedt}. We also use the abbreviations $\int_p\equiv\int \!\!
\frac{d^dp}{(2\pi)^d}$ and $v_d=1/(2^{d+1}\pi^{d/2} \Gamma(d/2))$,
such that $v_4=1/(32 \pi^2)$. Then, the threshold functions read

\begin{eqnarray*}
l_0^{(\mathrm{B/F/L/gh})d}(\omega) =&&\mkern-18mu\frac{k^{-d}}{4 v_d}\int_p \tilde\partial_t\log\left(P_{\mathrm{B}/\mathrm{F}/\mathrm{L}/\mathrm{gh}}+\omega k^2\right)  \\
l_{0 \mathrm{T/L}}^{(\mathrm{G})d}(\omega) =&&\mkern-18mu\frac{k^{-d}}{4 v_d}\int_p \tilde\partial_t\log\left(P_{\mathrm{GT}/\mathrm{GL}}+\omega k^2\right)  \\
l_{n_1,n_2}^{(\mathrm{FB})d}(\omega_1, \omega_2)&&\mkern-18mu =-\frac{k^{2(n_1+n_2)-d}}{4 v_d}\\
\times \int_p \tilde\partial_t &&\mkern-18mu \frac{1}{(P_\mathrm F + \omega_1 k^2)^{n_1}(P_\mathrm B + \omega_2 k^2)^{n_2}} \\
%
%
l_{n_1,n_2}^{(\mathrm{BG})d}(\omega_1, \omega_2)&&\mkern-18mu= -\frac{k^{2(n_1+n_2)-d}}{4 v_d} \\
\times\int_p \tilde\partial_t &&\mkern-18mu\frac{1}{(P_\mathrm B + \omega_1 k^2)^{n_1}(P_\mathrm{GT} + \omega_2 k^2)^{n_2}}
\end{eqnarray*}
\begin{eqnarray*}
m_2^{(\mathrm F)d}(\omega) =&&\mkern-18mu -\frac{k^{6-d}}{4 v_d} \int_p p^2 \tilde\partial_t \left( \frac{\partial}{\partial p^2}\frac{1}{P_\mathrm F + \omega k^2} \right)^2 \\
m_4^{(\mathrm F)d}(\omega) =&&\mkern-18mu -\frac{k^{4-d}}{4 v_d}\\
\times\int_p p^4 \tilde\partial_t &&\mkern-18mu  \left( \frac{\partial}{\partial p^2} \frac{1+r_{\mathrm L}}{P_\mathrm F + \omega k^2} \right)
\left( \frac{\partial}{\partial p^2} \frac{1+r_{\mathrm R}}{P_\mathrm F + \omega k^2} \right) \\
m_2^{(\mathrm{G})d}(\omega) =&&\mkern-18mu -\frac{k^{6-d}}{4v_d} \int_p p^2 \tilde\partial_t \left( \frac{\partial}{\partial p^2} \frac{1}{P_\mathrm{GT} + \omega k^2} \right)^2 
\end{eqnarray*}
\begin{eqnarray*}
m_{2,2}^{(\mathrm{B})d}(\omega_1, \omega_2) &&\mkern-18mu =-\frac{k^{6-d}}{4 v_d} \\
\times\int_p p^2 \tilde\partial_t &&\mkern-18mu\left( \frac{\tfrac{\partial}{\partial p^2}P_\mathrm B}{(P_\mathrm B + \omega_1 k^2)^2} \frac{\tfrac{\partial}{\partial p^2}P_\mathrm B}{(P_\mathrm B + \omega_2 k^2)^2} \right) \\
m_{1,2}^{(\mathrm{LB})d}(\omega_1,\omega_2)&&\mkern-18mu =-\frac{k^{4-d}}{4 v_d} \\
\times\int_p p^2 \tilde\partial_t &&\mkern-18mu \left( \frac{1+r_{\mathrm R}}{P_\mathrm F +\omega_1 k^2} \frac{\tfrac{\partial}{\partial p^2}P_\mathrm B}{(P_\mathrm B + \omega_2 k^2)^2} \right) \\
m_{1,2}^{(\mathrm{RB})d}(\omega_1,\omega_2)&&\mkern-18mu =-\frac{k^{4-d}}{4 v_d} \\
\times\int_p p^2 \tilde\partial_t &&\mkern-18mu \left( \frac{1+r_{\mathrm L}}{P_\mathrm F +\omega_1 k^2} \frac{\tfrac{\partial}{\partial p^2}P_\mathrm B}{(P_\mathrm B + \omega_2 k^2)^2} \right) \\
m_{1,2}^{(\mathrm{LG})d}(\omega_1,\omega_2)&&\mkern-18mu =-\frac{k^{4-d}}{4 v_d} \\
\times\int_p p^2 \tilde\partial_t &&\mkern-18mu\left( \frac{1+r_{\mathrm R}}{P_\mathrm F +\omega_1 k^2} \frac{\tfrac{\partial}{\partial p^2}P_\mathrm{GT}}{(P_\mathrm{GT} + \omega_2 k^2)^2} \right)
\end{eqnarray*}
\begin{eqnarray*}
a_1^d(\omega) =&&\mkern-18mu -\frac{k^{6-d}}{16v_d} \int_p \frac{1}{p^2}\tilde\partial_t \left( \frac{1}{P_\mathrm{GT} + \omega k^2} \right)^2 \\
a_3^d(\omega_1, \omega_2) = -&&\mkern-18mu \frac{k^{4-d}}{4v_d} \int_p \tilde\partial_t \left( \frac{1+r_{\mathrm R}}{P_\mathrm F +\omega_1 k^2} \frac{1}{P_\mathrm{GT} + \omega_2 k^2} \right) \, .
\end{eqnarray*}
%

For practical computations, we use the linear regulator for the scalar
bosons, for the gauge bosons and for the ghosts
\begin{equation}\label{eq:cutoff}
x r_{\mathrm{B}/\mathrm{GT}/\mathrm{GL}/\mathrm{gh}}(x)=(1-x)\theta(1-x),
\end{equation}
where $x=q^2/k^2$. For the spinor fermions the linear regulator corresponds to a shape function
$r_{\mathrm{L}/\mathrm{R}}$ such that $x(1+r_{\mathrm{B}}(x))=x(1+r_{\mathrm{L}/\mathrm{R}}(x))^2$.  
This regulator satisfies an optimization criterion within our present truncation and is
technically advantageous, as we can perform all momentum integrations
analytically, obtaining
\begin{eqnarray*}
l_0^{(\mathrm{B})d}(\omega) &=& \frac{2}{d}\frac{1-\tfrac{\eta_\phi}{d+2} }{1+\omega} \, , \\
l_0^{(\mathrm{F})d}(\omega) &=& \frac{2}{d}\frac{1-\tfrac{\eta_\mathrm{L}+\eta_\mathrm{R}}{2(d+1)}}{1+\omega} \, , \\
l_0^{(\mathrm{L})d}(\omega) &=& \frac{2}{d}\frac{1-\tfrac{\eta_\mathrm{L}}{d+1}}{1+\omega} \, , \\
l_{0\mathrm T}^{(\mathrm{G})d}(\omega) &=& \frac{2}{d}\frac{1-\tfrac{\eta_\mathrm W}{d+2}}{1+\omega} \, , \\
l_{0\mathrm L}^{(\mathrm{G})d}(\omega) &=& \frac{2}{d}\frac{1-\tfrac{\eta_\phi}{d+2}}{1+\omega} \, , \\
l_0^{(\mathrm{gh})d}(\omega) &=& \frac{2}{d}\frac{1}{1+\omega} \, ,
\end{eqnarray*}
\begin{eqnarray*}
l_{n_1,n_2}^{(\mathrm{FB})d}(\omega_1, \omega_2) &=& \frac{2}{d}\Bigg[
n_1\frac{ 1-\frac{\eta_\mathrm{L}+\eta_\mathrm{R}}{2(d+1)} }{ (1+\omega_1)^{1+n_1}(1+\omega_2)^{n_2} } \\
&&+\;\, n_2\frac{ 1-\frac{\eta_\phi}{d+2} }{ (1+\omega_1)^{n_1}(1+\omega_2)^{1+n_2} } \Bigg] \, , \\
%
%
l_{n_1,n_2}^{(\mathrm{BG})d}(\omega_1, \omega_2) &=& \frac{2}{d}\Bigg[
n_1\frac{1-\frac{\eta_\phi}{d+2}}{(1+\omega_1)^{1+n_1}(1+\omega_2)^{n_2}} \\
&&+\;\, n_2\frac{1-\frac{\eta_\mathrm W}{d+2}}{(1+\omega_1)^{n_1}(1+\omega_2)^{1+n_2}}\Bigg] \, ,
\end{eqnarray*}
\begin{eqnarray*}
m_2^{(\mathrm F)d}(\omega) &=& \frac{1}{(1 + \omega)^4} \, , \\
m_4^{(\mathrm F)d}(\omega) &=& \frac{1}{(1 + \omega)^4} + 
\frac{1-\tfrac{1}{2}(\eta_\mathrm L + \eta_\mathrm R)}{(d-2)(1+\omega)^3} \\
&-&\left( \frac{1-\tfrac{1}{2}(\eta_\mathrm L + \eta_\mathrm R)}{2d-4} + \frac{1}{4} \right)\frac{1}{(1 + \omega)^2} \, , \\
m_2^{(\mathrm{G})d}(\omega) &=& \frac{1}{(1 + \omega)^4} \, ,
\end{eqnarray*}
\begin{eqnarray*}
m_{2,2}^{(\mathrm{B})d}(\omega_1, \omega_2) &=& \frac{1}{(1+\omega_1)^2(1+\omega_2)^2} \, , \\
m_{1,2}^{(\mathrm{LB})d}(\omega_1,\omega_2) &=& \frac{1-\tfrac{\eta_\phi}{d+1}}{(1+\omega_1)(1+\omega_2)^2} \, , \\
m_{1,2}^{(\mathrm{RB})d}(\omega_1,\omega_2) &=& \frac{1-\tfrac{\eta_\phi}{d+1}}{(1+\omega_1)(1+\omega_2)^2} \, , \\
m_{1,2}^{(\mathrm{LG})d}(\omega_1,\omega_2) &=& \frac{1-\tfrac{\eta_\mathrm W}{d+1}}{(1+\omega_1)(1+\omega_2)^2} \, ,
\end{eqnarray*}
\begin{eqnarray*}
a_1^d(\omega) &=& \frac{1- \tfrac{\eta_\mathrm W}{d}}{d-2}\frac{1}{(1 + \omega)^3} \, , \\
a_3^d(\omega_1, \omega_2) &=& \frac{2}{d-1}\frac{1-\tfrac{\eta_\mathrm W}{d+1}}{(1+\omega_1)(1+\omega_2)^2} \\
&+& \frac{1}{d-1} \frac{\left( 1-\tfrac{\eta_\mathrm L}{d} \right) - \omega_1\left( 1-\tfrac{\eta_\mathrm R}{d} \right)}{(1+\omega_1)^2(1+\omega_2)} \, .
\end{eqnarray*}

\section{Derivation of the flow equations for the matter sector}\label{sec:deriv-matter-flow}

The computation of the RG flow of the matter sector inside the truncation \eqref{eq:truncation} will be sketched.
For further details see \cite{thesis}.

\subsection{Flow equation for the potential}

The flow of the potential can be computed by setting the field
$\phi^a$ to a constant value and all the other fields to zero. This
projects both sides of the flow onto the scalar potential.  Then, in
Landau gauge the matrix $\Gamma_k^{(2)}+R_k$ can be inverted easily.
Multiplying with the derivative of the regulator, and taking the
supertrace yields the result.  This can be interpreted as an improved
one-loop computation for a $0$-point function, i.e. a sum over all the
one-loop graphs with no external legs.  The gauge contribution takes
the form of a closed gauge boson propagator, and since it does not
involve any vertex, it should not explicitly depend on
$\bar{g}$. Indeed we get
\begin{widetext}
\begin{eqnarray*}
\partial_tU=&&\frac{1}{2}\int \frac{d^dp}{(2\pi)^d}\partial_t P_\mathrm B\left[ \frac{2\NL-1}{Z_\phi P_\mathrm B + U'}+\frac{1}{Z_\phi P_\mathrm B + U' + 2\rho U''} \right] \\
\\
&&-d_{\gamma} \Ng\int\frac{d^dp}{(2\pi)^d}\left\{ \left[(\NL-1)+\frac{Z_\mathrm L Z_\mathrm R P_\mathrm F}{Z_\mathrm L Z_\mathrm R P_\mathrm F + \rho \bar{h}^2}\right]\frac{\partial_t[Z_\mathrm L r_{\mathrm L}]}{Z_\mathrm L (1+r_{\mathrm L})} + \frac{Z_\mathrm LZ_\mathrm R P_\mathrm F}{Z_\mathrm L Z_\mathrm R P_\mathrm F + \rho \bar{h}^2}\frac{\partial_t[Z_\mathrm R r_{\mathrm R}]}{Z_\mathrm R (1+r_{\mathrm R})}\right\} \\
\\
&& + \frac{1}{2}\sum_{i=1}^{\NL^2-1}\int\frac{d^dp}{(2\pi)^d}\left[ (d-1)\frac{\partial_t(Z_W p^2r_{\mathrm{GT}})}{Z_W P_\mathrm{GT} + \bar{m}_{W,i}^2} +\frac{\partial_t(Z_\phi p^2r_{\mathrm{GL}})}{Z_\phi P_\mathrm{GL}} \right] - \int\frac{d^dp}{(2\pi)^d}\frac{\dG p^2\partial_t r_{\mathrm{gh}}}{P_{\mathrm{gh}}}
\end{eqnarray*}
that is, in terms of threshold functions
\begin{eqnarray*}
\partial_t U = 2 v_d k^d \Big\{ (2\NL-1)l_0^{(\mathrm B)d}\left( \tfrac{U'}{Z_\phi k^2} \right) 
+ l_0^{(\mathrm B)d}\left( \tfrac{U' + 2 \rho U''}{Z_\phi k^2} \right) - d_\gamma \Ng \left[ 
(\NL-1)l_{0}^{(\mathrm L)d}\left( 0 \right) + 2 l_{0}^{(\mathrm F)d}\left( \tfrac{\rho \bar{h}^2}{Z_\mathrm L Z_\mathrm R k^2} \right) 
\right] \\
- 2\dG l_0^{(\mathrm{gh})d}\left( 0 \right) + \sum_{i=1}^{\dG} \left[ (d-1)l_{0\mathrm T}^{(\mathrm{G})d}\left( \tfrac{\bar{m}_{W,i}^2}{Z_W k^2} \right) + l_{0\mathrm L}^{(\mathrm{G})d}\left( 0 \right) \right] \Big\}
\end{eqnarray*}
where $U$ is a function of $\rho$. Switching to dimensionless
quantities this becomes Eq.~\eqref{floweq:potential}.
\end{widetext}
%

\subsection{Flow equation for the Yukawa coupling}\label{subsection:yukawaflow}

For the derivation of the flow of the Yukawa coupling, we first
separate the bosonic field into the vev and a purely radial deviation
from the vev. This corresponds to setting $\Delta\phi_2=0$ in
\Eqref{fluctandvev}. While this is irrelevant in the symmetric regime,
it makes a difference in the SSB regime, as it projects onto the the
Yukawa coupling between the fermions and Higgs boson, being the radial mode.  The
projection of the flow equation onto such an operator reads
\begin{equation}\label{eq:flowh}
\partial_t\bar{h}=-\frac{1}{\sqrt{2}}\frac{\overrightarrow{\delta}}{\delta\bar{\psi}_{\mathrm{L}}^{\hat{n}}(p)}\frac{\overrightarrow{\delta}}{\delta\Delta\phi_1^{\hat{n}}(p')} \partial_t\Gamma_k \frac{\overleftarrow{\delta}}{\delta\psi_{\mathrm{R}}(q)}
\Bigg|_0\,.
\end{equation}
The vertical line indicates that the equation is evaluated at
vanishing momenta $p'=p=q=0$ and at vanishing fluctuation fields.
Next, we can decompose the matrix $(\Gamma_k^{(2)}+R_k)$ into two
parts.  One part, which we call $(\Gamma_{k,0}^{(2)}+R_k)$, contains
only $v$ and is independent of the fluctuations.  The remaining part,
$\Delta \Gamma_k^{(2)}$, contains all fluctuating fields. Using the $\tilde \partial_t$-notation of App.~\ref{section:threshold} and expanding by means of the Mercator series, 
the flow equation can be written as
\begin{eqnarray}\label{eq:logarithm}
\partial_t \Gamma_k&=&\frac{1}{2}\sum_{s=1}^{\infty}\frac{(-)^{s+1}}{s}{\rm STr}\tilde \partial_t \left[
\left( \frac{\Delta \Gamma_k^{(2)}}{\Gamma_{k,0}^{(2)}+R_k} \right)^s\right] \nonumber\\
&+&\frac{1}{2}{\rm STr}\tilde \partial_t \log(\Gamma_{k,0}^{(2)}+R_k)\,.
\end{eqnarray}
Plugging this expression into equation \eqref{eq:flowh}, only the term
to third power in $\Delta\Gamma_k^{(2)}$ survives the projection.
Since we took three derivatives of the Wetterich equation, the
diagrammatic interpretation of the result is in terms of one-loop
graphs with three external legs: two fermions of opposite chirality
and one radial scalar.  The gauge contribution comes from triangular
loops with three different propagators: one scalar, one spinor and one
gauge vector.  It always involves the two-scalars-one-vector vertex.
This vertex is proportional to the difference of incoming scalar
momenta, while the gauge boson propagator in Landau gauge is
transverse.  These two facts plus conservation of momentum entail that
the direct gauge contribution to the momentum-independent Yukawa
coupling under consideration vanishes in our truncation. This formal
argument can straightforwardly be verified by performing the matrix
calculations and taking the supertrace, yielding
\begin{widetext}
\begin{eqnarray*}
\partial_t \bar{h} = - \frac{\bar{h}^3}{2}\int\frac{d^dp}{(2\pi)^d}\tilde{\partial}_t \left[ \frac{1}{Z_\mathrm L Z_\mathrm R P_\mathrm F + \rho \bar{h}^2}
\left(\frac{2 \rho U''}{(Z_\phi P_\mathrm B + U')^2} - \frac{ 6\rho U'' + 4\rho^2 U'''}{(Z_\phi P_\mathrm B + U' + 2\rho U'')^2}\right)\right. \\
\\
+\frac{2\rho \bar{h}^2}{(Z_\mathrm L Z_\mathrm R P_\mathrm F + \rho \bar{h}^2)^2} \left(\frac{1}{Z_\phi P_\mathrm B + U'} - \frac{1}{Z_\phi P_\mathrm B + U' + 2\rho U''}\right)\\
\\
\left. - \frac{1}{Z_\mathrm L Z_\mathrm R P_\mathrm F + \rho \bar{h}^2} \left(\frac{1}{Z_\phi P_\mathrm B + U'} - \frac{1}{Z_\phi P_\mathrm B + U' + 2\rho U''}\right)\right]
\end{eqnarray*}
where the whole r.h.s. should be evaluated at the value $\rho=\frac{1}{2}{\bar v}^2$ that minimizes the potential $U$.
In terms of the threshold functions as defined in App. \ref{section:threshold} this reads
\begin{align*}
\partial_t \bar{h}^2	 = \frac{4v_d \bar{h}^4}{Z_\mathrm L Z_\mathrm R Z_\phi k^{4-d}} \Big[ 
\frac{2\rho U''}{Z_\phi k^2} 
&l_{1,2}^{(\mathrm{FB})d}\left(\tfrac{\rho\bar{h}^2}{Z_\mathrm L Z_\mathrm R k^2},\tfrac{U'}{Z_\phi k^2}\right) 
- \frac{6\rho U''+4\rho U'''}{Z_\phi k^2} 
l_{1,2}^{(\mathrm{FB})d}\left( \tfrac{\rho \bar{h}^2}{Z_\mathrm L Z_\mathrm R k^2}, \tfrac{U'+2\rho U''}{Z_\phi k^2} \right) \\
+ \frac{2\rho\bar{h}^2}{k^2} 
&l_{2,1}^{(\mathrm{FB})d}\left( \tfrac{\rho \bar{h}^2}{Z_\mathrm L Z_\mathrm R k^2}, \tfrac{U'}{Z_\phi k^2} \right) 
- \frac{2\rho\bar{h}^2}{k^2}  
l_{2,1}^{(\mathrm{FB})d}\left( \tfrac{\rho \bar{h}^2}{Z_\mathrm L Z_\mathrm R k^2}, \tfrac{U'+2\rho U''}{Z_\phi k^2} \right) \\
- &l_{1,1}^{(\mathrm{FB})d}\left( \tfrac{\rho \bar{h}^2}{Z_\mathrm L Z_\mathrm R k^2}, \tfrac{U'}{Z_\phi k^2} \right) + 
l_{1,1}^{(\mathrm{FB})d}\left( \tfrac{\rho \bar{h}^2}{Z_\mathrm L Z_\mathrm R k^2}, \tfrac{U'+2\rho U''}{Z_\phi k^2} \right) \Big]\, .
\end{align*}
Switching over to dimensionless quantities, we end up with the representation \eqref{floweq:yukawa} given in the main text.

\end{widetext}
%
\subsection{Flow of the scalar anomalous dimension}

For the derivation of the flow of $Z_{\phi }$, we decompose the bosonic field as in App. \ref{subsection:yukawaflow}. The projection of the Wetterich equation onto the massive scalar kinetic term leads us to
\begin{eqnarray} \label{eq:etaphiflow}
 \partial_t Z_{\phi }&=&-\frac{\partial}{\partial(p'^2)}\frac{\delta}{\delta \Delta\phi_1^{\hat n}(p')}
\frac{\delta}{\delta \Delta\phi_1^{\hat n}(q')} \partial_t\Gamma_k \Bigg|_0\,.\nonumber
\end{eqnarray}
As before the vertical line indicates that the equation is evaluated
at vanishing momenta $p'=q'=0$ and at vanishing fluctuation fields.
Expanding again the r.h.s. of the flow equation according to
eq.~\eqref{eq:logarithm}, this time only the second order term ($s=2$)
contributes.  Since two derivatives of the flow equation have to be
taken, the result can diagrammatically be interpreted as one-loop
graphs with two external scalar legs. From a one-loop analysis we
expect two kinds of gauge contributions.  One is due to the
two-scalars-one-vector vertex and produces a loop containing one
scalar and one gauge boson propagator.  This is present in both the
symmetric and in the spontaneously broken regimes.  Another is due to
the two-scalars-two-vectors vertex. If two external scalar legs are
identified with the vev, the corresponding loop contains two gauge
boson propagators. Therefore this contribution will be present only in
the SSB regime.  Indeed, performing the matrix calculations and taking
the supertrace we find
\begin{widetext}
\begin{eqnarray*}
\partial_t Z_\phi &=& \frac{1}{d}\int\frac{d^dp}{(2\pi)^d}\tilde{\partial}_t \\
&&\left\{ \left[ (3\sqrt{2\rho}U''+2\sqrt{2\rho^3}U''')^2p^2Z^2_\phi \left( \frac{\tfrac{\partial}{\partial p^2}P_\mathrm B}{(Z_\phi P_\mathrm B+U'+2\rho U'')^2} \right)^2 
+(2\NL-1)2\rho U''^2p^2Z^2_\phi \left( \frac{\tfrac{\partial}{\partial p^2}P_\mathrm B}{(Z_\phi P_\mathrm B+U')^2} \right)^2 \right] \right. \\
&&+ d_{\gamma}\Ng\bar{h}^2\left[ 2p^4Z_\mathrm{L} Z_\mathrm{R} \left( \frac{\partial}{\partial p^2}\frac{1+r_{\mathrm L}}{Z_\mathrm L Z_\mathrm R P_\mathrm F+\rho \bar{h}^2} \right)
\left( \frac{\partial}{\partial p^2}\frac{1+r_{\mathrm R}}{Z_\mathrm L Z_\mathrm R P_\mathrm F+\rho \bar{h}^2} \right)
-2\rho \bar{h}^2p^2\left( \frac{\partial}{\partial p^2}\frac{1}{Z_\mathrm L Z_\mathrm R P_\mathrm F+\rho \bar{h}^2} \right)^2 \right]\\
&&-4(d-1) \bar{g}^2  Z^2_\phi \sum_{a=1}^{\NL}\sum_{i=1}^{\dG}\frac {T^i_{{\hat n}a}T^i_{a{\hat n}}}{(Z_\phi P_\mathrm B + U')(Z_W P_\mathrm{GT} + \bar{m}_{W,i}^2)} \\
&&\left. + \frac{(d-1)}{\rho}\sum_{i=1}^{\dG} \bar{m}_{W,i}^4 \left( \frac{1}{p^2(Z_W P_\mathrm{GT} + \bar{m}_{W,i}^2)^2} 
+ 2p^2\left( \frac{\partial}{\partial p^2}\frac{1}{Z_W P_\mathrm{GT} + \bar{m}_{W,i}^2} \right)^2 \right)\right\}. 
\end{eqnarray*}
Again the whole r.h.s. should be evaluated at the value $\rho=\frac{1}{2}\bar{v}^2$ that minimizes the potential $U$.
Translating this result in terms of threshold functions, yields
\begin{eqnarray*}
\partial_t Z_\phi &=& -\frac{8v_d}{Z_\phi^2 k^{6-d}d}\left[ 
\left(3\sqrt{\rho}U''+2\sqrt{\rho^3}U'''\right)^2 
m_{2,2}^{(\mathrm B)d}\left(\tfrac{U'+2\rho U''}{Z_\phi k^2},\tfrac{U'+2\rho U''}{Z_\phi k^2} \right) 
+(2\NL-1)\rho U''^2 
m_{2,2}^{(\mathrm B)d}\left(\tfrac{U'}{Z_\phi k^2},\tfrac{U'}{Z_\phi k^2}\right) \right] \\
&& -\frac{4v_d d_\gamma\Ng}{d} \left[
\frac{2\bar{h}^2}{Z_\mathrm L Z_\mathrm R k^{4-d}} 
m_4^{(\mathrm F)d}\left(\tfrac{\rho \bar{h}^2}{Z_\mathrm L Z_\mathrm R k^2}\right) 
- \frac{2\rho \bar{h}^4}{Z_\mathrm L^2 Z_\mathrm R^2k^{6-d}} 
m_2^{(\mathrm F)d}\left( \tfrac{\rho \bar{h}^2}{Z_\mathrm L Z_\mathrm R k^2} \right) \right] \\
&&+\frac{16 v_d (d-1)}{d}  \frac{\bar{g}^2 Z_\phi}{k^{4-d}Z_W}  
\sum_{a=1}^{\NL}\sum_{i=1}^{\dG} T^i_{{\hat n}a}T^i_{a{\hat n}} \, 
l_{1,1}^{(\mathrm{BG})d}\left( \tfrac{U'}{Z_\phi k^2}, \tfrac{\bar{m}_{W,i}^2}{Z_W k^2} \right) \\
&& -\frac{8v_d (d-1)}{d}
\sum_{i=1}^{\dG} \frac{\bar{m}_{W,i}^4}{Z_W^2 k^{6-d}\rho} \left[
2 a_1^d\left( \tfrac{\bar{m}_{W,i}^2}{Z_W k^2} \right) 
+ m_2^{(\mathrm{G})d}\left( \tfrac{\bar{m}_{W,i}^2}{Z_W k^2} \right) \right].
\end{eqnarray*}
In terms of dimensionless quantities, this leads to Eq.~\eqref{floweq:etaphi}.

\end{widetext}
%
\subsection{Flow of the spinor anomalous dimensions}

For the anomalous dimensions of the spinors, the procedure is very
similar to the one explained for the scalar. In the broken regime, the
wave function renormalization of the left-handed spinors in principle
splits into two functions. Here we concentrate only on the wave
function renormalizations associated with the massive top quark,
i.e. those for the ${\hat n}$-th left-handed component and for the right-handed one. We start
with the projection
\begin{equation}
\partial_tZ_{\mathrm{L/R},k}=\left.\frac{-1}{2v_dd_{\gamma} \Ng}\mathrm{tr}\gamma^{\mu}\frac{\partial}{\partial p'^{\mu}}\frac{\overrightarrow{\delta}}{\delta\bar{\psi}_{\mathrm{L/R}}^{\hat n}(p')}\partial_t \Gamma_k
\frac{\overleftarrow{\delta}}{\delta\psi_{\mathrm{L/R}}^{\hat
    n}(q')}\right|_0 \nonumber
\end{equation}
where the trace is over spinor and generation indices.
As before, the vertical line denotes that the equation is evaluated at
vanishing momenta $p'=q'=0$ and at vanishing fluctuation fields.
Expanding the r.h.s. of the flow equation according to
Eq.~\eqref{eq:logarithm}, only the second order term ($s=2$)
contributes.  Obviously, the right-handed fermion does not receive
direct corrections from the gauge boson whereas the left-handed
fermion does. The gauge-field-independent contributions differ from the
results of~\cite{Gies:2009sv} by a factor of two due
to a qualitatively irrelevant prefactor error in the earlier paper.
For the right-handed spinor, the result is
\begin{widetext}
\begin{eqnarray*}
\partial_t Z_\mathrm R = \frac{\bar{h}^2}{d}\int\!\! \frac{d^dp}{(2\pi)^d}\, p^2\tilde{\partial}_t \Bigg[ \frac{Z_\mathrm R (1+r_{\mathrm R})}{Z_\mathrm L Z_\mathrm R P_\mathrm F+\rho \bar{h}^2}\left( \frac{Z_\phi \tfrac{\partial}{\partial p^2}P_\mathrm B}{(Z_\phi P_\mathrm B+U'+2\rho U'')^2}+\frac{Z_\phi \tfrac{\partial}{\partial p^2}P_\mathrm B}{(Z_\phi P_\mathrm B+U')^2} \right)\\
+ 2(\NL-1)\frac{Z_\mathrm R(1+r_\mathrm{R})}{Z_\mathrm L Z_\mathrm R P_\mathrm F}\frac{Z_\phi \tfrac{\partial}{\partial p^2}P_\mathrm B}{(Z_\phi P_\mathrm B+U')^2} \Bigg]\,.
\end{eqnarray*}
In terms of threshold functions, this reads
\begin{equation*}
\partial_t Z_\mathrm R = -\frac{4v_d \bar{h}^2}{d Z_\phi Z_\mathrm L k^{4-d}} \left[ m_{1,2}^{(\mathrm{LB})d}(\tfrac{\rho \bar{h}^2}{Z_\mathrm L Z_\mathrm R k^2}, \tfrac{U' + 2\rho U''}{Z_\phi k^2}) 
+ m_{1,2}^{(\mathrm{LB})d}(\tfrac{\rho \bar{h}^2}{Z_\mathrm L Z_\mathrm R k^2}, \tfrac{U'}{Z_\phi k^2}) + 2(\NL-1)m_{1,2}^{(\mathrm{LB})d}(0, \tfrac{U'}{Z_\phi k^2}) \right],
\end{equation*}
which leads to Eq.~\eqref{floweq:etaR} in terms of dimensionless
quantities. 
For the left-handed fermion, the result is
\begin{eqnarray*}
\partial_t Z_\mathrm L \!\!&=&\!\! \int\!\! \frac{d^dp}{(2\pi)^d}\,\tilde{\partial}_t \left\{ \frac{\bar{h}^2}{d}\frac{Z_\mathrm L p^2(1+r_{\mathrm L})}{Z_\mathrm L Z_\mathrm R P_\mathrm F+\rho \bar{h}^2} \left( \frac{Z_\phi \tfrac{\partial}{\partial p^2}P_\mathrm B}{(Z_\phi P_\mathrm B+U'+2\rho U'')^2}+\frac{Z_\phi \tfrac{\partial}{\partial p^2}P_\mathrm B}{(Z_\phi P_\mathrm B+U')^2} \right) \right.\\
&-&\!\!\frac{(d-1)}{d}\bar{g}^2 Z^2_\mathrm L \sum_{i=1}^{\dG} (T^i_{{\hat n}{\hat n}})^2 \left[ \left( \frac{2}{Z_W P_\mathrm{GT} + \bar{m}_{W,i}^2} 
+ 2p^2\frac{\partial}{\partial p^2}\frac{1}{Z_W P_\mathrm{GT} + \bar{m}_{W,i}^2} \right) 
\left(\frac{Z_\mathrm R(1 + r_\mathrm R)}{Z_\mathrm L Z_\mathrm R P_\mathrm F + \rho \bar{h}^2}-
\frac{Z_\mathrm R(1 + r_\mathrm R)}{Z_\mathrm L Z_\mathrm R P_\mathrm F }\right)
 \right]\\
&-&\!\!\left.\frac{(d-1)}{d}\bar{g}^2 Z^2_\mathrm L \sum_{a=1}^{\NL}\sum_{i=1}^{\dG} T^i_{{\hat n}a}T^i_{a{\hat n}} \left[ \left( \frac{2}{Z_W P_\mathrm{GT} + \bar{m}_{W,i}^2} 
+ 2p^2\frac{\partial}{\partial p^2}\frac{1}{Z_W P_\mathrm{GT} + \bar{m}_{W,i}^2} \right) \frac{Z_\mathrm R(1 + r_\mathrm R)}{Z_\mathrm L Z_\mathrm R P_\mathrm F} \right]\right\}.
\end{eqnarray*}
In terms of threshold functions, we obtain
\begin{eqnarray*}
\partial_t Z_\mathrm L &=&-\frac{4v_d \bar{h}^2}{d Z_\phi Z_\mathrm R k^{4-d}} \left[
m_{1,2}^{(\mathrm{RB})d}\left(\tfrac{\rho h^2}{Z_\mathrm L Z_\mathrm R k^2}, \tfrac{U' + 2\rho U''}{Z_\phi k^2}\right) 
+m_{1,2}^{(\mathrm{RB})d}\left(\tfrac{\rho h^2}{Z_\mathrm L Z_\mathrm R k^2}, \tfrac{U'}{Z_\phi k^2}\right) \right] 
-\frac{8 v_d(d-1)}{d}\frac{\bar{g}^2 Z_\mathrm L}{Z_W k^{4-d}}\Bigg\{\\
&& \sum_{i=1}^{\dG} (T^i_{{\hat n}{\hat n}})^2 \left[ 
m_{1,2}^{(\mathrm{LG})d}\left(\tfrac{\rho h^2}{Z_\mathrm L Z_\mathrm R k^2}, \tfrac{\bar{m}_{W,i}^2}{Z_W k^2} \right)
-m_{1,2}^{(\mathrm{LG})d}\left(0, \tfrac{\bar{m}_{W,i}^2}{Z_W k^2} \right)
- a_3^d\left(\tfrac{\rho h^2}{Z_\mathrm L Z_\mathrm R k^2}, \tfrac{\bar{m}_{W,i}^2}{Z_W k^2} \right) 
+ a_3^d\left(0, \tfrac{\bar{m}_{W,i}^2}{Z_W k^2} \right)\right]\\
&&+\sum_{a=1}^{\NL}\sum_{i=1}^{\dG} T^i_{{\hat n}a}T^i_{a{\hat n}} \left[ m_{1,2}^{(\mathrm{LG})d}\left(0, \tfrac{\bar{m}_{W,i}^2}{Z_W k^2} \right) - a_3^d\left(0, \tfrac{\bar{m}_{W,i}^2}{Z_W k^2} \right) \right]\Bigg\},
\end{eqnarray*}
the translation into dimensionless quantities of which agrees with Eq.~\eqref{floweq:etaLgeneral}.
Upon a global color rotation, we can choose (without loss of
generality) the direction of the vev ${\hat n}$ to point along a
single color axis,
i.e. ${\hat n}^a\propto \delta^{aA}$. Then, this anomalous dimension takes a simpler form, given in Eq.~\eqref{floweq:etaLsimple}.

\end{widetext}
%

\section{Flow equation for the gauge coupling}\label{sec:deriv-gauge-flow}

In this appendix, we set the spacetime dimension to $d=4$, and we focus on the
gauge group SU$(\NL)$.
Since we will be satisfied with the one loop beta function we set all
the wave function renormalizations to one (terms of order ${\cal
  O}(\partial_t Z)$ on the r.h.s. of the flow equation lead to higher
loop corrections of the $\beta_{g^2}$ function).  Still, relevant
nonperturbative information will arise from the threshold behavior
describing the decoupling of massive modes.

\subsection{Contribution from the gauge modes}
\label{sec:appgauge}

For the gauge contribution, the relevant part of the effective Lagrangian is
\begin{eqnarray*}
{\cal L}_k&\ni&\frac{1}{4}
(F^i_{\mu\nu})^2
+\bar{g}^2 W^i_\mu W^j_\mu \phi^{\dagger a}T^i_{ab}T^j_{bc}\phi^c
+{\cal L}_{k,\text{gf}}+{\cal L}_{k,\text{gh}}\\
 &=&\frac{1}{4}
(F^i_{\mu\nu})^2
+\frac{\bar{g}^2 \bar{v}^2}{2} \hat{n}^{\dagger a}T^i_{ab}T^j_{bc}\hat{n}^c W^i_\mu W^j_\mu
+\cdots .
\end{eqnarray*}
This defines the mass matrix for the gauge bosons, as given in
eq.~\eqref{gaugemass}.  As the generators are real, the mass matrix
has real eigenvalues. In order to compute the running coupling, we use
the background-field method and project on the operator $F^2/4$.  To the present one-loop
order, no distinction between the background field and the fluctuation
field has to be made \cite{Reuter:1993kw,Pawlowski:2001df}, such 
that it suffices to compute the Hessian $\bar{\Gamma}_k^{(2)}$ at zero
fluctuation field. Here and in the following, we use the notation of
\cite{Reuter:1993kw,Gies:2002af}.  This Hessian for the $W$-boson reads
\be
\bar{\Gamma}^{(2)\, ij}_{k\ \ \mu\nu}\Big|_W={\cal D}_{\mathrm T\,
  \mu\nu}^{\phantom{T}\,
  ij}+\left(1-\frac{1}{\alpha}\right)D_\mu^{il}D_\nu^{lj}\nonumber 
+\bar{m}^{2\ \ ij}_W\delta_{\mu\nu},
\ee
where ${\cal D}_{\mathrm T\, \mu\nu} = -D^2 \delta_{\mu\nu} + 2 i \bar{g} F_{\mu\nu}$.
The contributions from ghost fluctuations are
\be
\bar{\Gamma}^{(2)\, ij}_k\Big|_{gh}=-D_\mu^{il}D_\mu^{lj}+O(\alpha^2) \equiv
-(D^2)^{ij}+O(\alpha^2).\nonumber
\ee
As we will focus on Landau gauge ($\alpha\rightarrow 0$), we ignore
from now on the ghost-Higgs contributions $\sim O(\alpha^2)$.  For a
covariantly-constant background field, projectors onto the
longitudinal and transverse subspaces w.r.t. the background field
exist,
\be
\Pi_{\text{T}}+\Pi_{\text{L}}=\mathds{1} \quad , \quad
\Pi_{\text{T}/\text{L}}^2=\Pi_{\text{T}/\text{L}} \quad , \quad
\Pi_{\text{T}} \Pi_{\text{L}}=0,\nonumber
\ee
with $\Pi_{\text{L} \mu\nu}=-(\mathcal{D}^{-1}_{\text{T}})_{\mu\lambda}
D_\lambda D_\nu$ and $\Pi_{\text{T}} = \mathds{1} - \Pi_{\text{L}}$, such that
\begin{align*}
\bar{\Gamma}^{(2)\, ij}_{k\ \ \mu\nu}\Big|_W=\ &\Pi_{\text{T}\, \mu\lambda}^{\phantom{T}\, il}\left[ {\cal D}_{\mathrm T\, \lambda\nu}^{\phantom{T}\, lj}+\bar{m}^{2\ \ ij}_W\delta_{\lambda\nu}\right]&\\
+\ &\Pi_{\text{L}\, \mu\lambda}^{\phantom{T}\, il}\left[\frac{1}{\alpha} {\cal D}_{\mathrm T\, \lambda\nu}^{\phantom{T}\, lj}+\bar{m}^{2\ \ ij}_W\delta_{\lambda\nu}\right]&\nonumber
\end{align*}
We choose a similar decomposition for the regulator
\be
R_k\Big|_W=\Pi_{\text{T}} {\cal D}_\mathrm T r_k\!\left(\frac{{\cal D}_\mathrm T}{k^2}\right)+\Pi_{\text{L}} \frac{1}{\alpha} {\cal D}_\mathrm T r_k\!\left(\frac{{\cal D}_\mathrm T}{k^2}\right)\nonumber
\ee
hence also the functional trace on the r.h.s. of the flow equation decomposes into these two sectors.
Using the important property that
\be
{\rm Tr}\left[\Pi_{\text{L}} f({\cal D}_\mathrm T)\right]={\rm Tr}\left[f(-D^2)\right],\nonumber
\ee
we get
\begin{align*}
{\rm Tr}\left[\frac{\partial_t{R}_k}{\Gamma_k^{(2)}+R_k}\right]_W=&
{\rm Tr}\left[\Pi_{\text{T}}\frac{{\cal D}_\mathrm T\partial_t{r}_k\!\left(\frac{{\cal D}_\mathrm T}{k^2}\right)}{{\cal D}_\mathrm T \left(1+r\!\left(\frac{{\cal D}_\mathrm T}{k^2}\right)\right)+\bar{m}^2_W}\right]&\\
+&{\rm Tr}\left[\frac{ (-D^2)\partial_t{r}_k\!\left(\frac{-D^2}{k^2}\right)}{(-D^2)\left(1+r\!\left(\frac{-D^2}{k^2}\right)\right)+\alpha \bar{m}^2_W}\right]&\
\end{align*}
and writing $\Pi_{\text{T}}=1-\Pi_{\text{L}}$ in the first term we
obtain two unconstrained traces for different differential operators.
The ghost contribution gives
\be
{\rm Tr}\left[\frac{\partial_t{R}_k}{\Gamma_k^{(2)}+R_k}\right]_{\text gh}=
-2{\rm Tr}\left[\frac{(-D^2)\partial_t{r}_k\!\left(\frac{-D^2}{k^2}\right)}{(-D^2)\left(1+r\!\left(\frac{-D^2}{k^2}\right)\right)}\right]\nonumber
\ee
such that the total contribution reads in the Landau gauge $\alpha\rightarrow 0$
\begin{widetext}
\begin{equation}
\label{totgaugetogauge}
{\rm STr}\!\left[\frac{\partial_t{R}_k}{\Gamma_k^{(2)}+R_k}\right]_{W}\!\!=\!\!=
{\rm Tr}\!\left[\frac{{\cal D}_\mathrm T\partial_t{r}_k\!\left(\frac{{\cal D}_\mathrm T}{k^2}\right)}{{\cal D}_\mathrm T \left(1+r\!\left(\frac{{\cal D}_\mathrm T}{k^2}\right)\right)+\bar{m}^2_W}\right]\!
-{\rm Tr}\!\left[\frac{(-D^2)\partial_t{r}_k\!\left(\frac{-D^2}{k^2}\right)}{(-D^2)\left(1+r\!\left(\frac{-D^2}{k^2}\right)\right)+\bar{m}^2_W}\right]\!
-{\rm Tr}\!\left[\frac{(-D^2)\partial_t{r}_k\!\left(\frac{-D^2}{k^2}\right)}{(-D^2)\left(1+r\!\left(\frac{-D^2}{k^2}\right)\right)}\right] .
\end{equation}
\end{widetext}
To simplify the calculation, we choose a basis in adjoint color space
where the gauge boson mass matrix is diagonal, as
in~\Eqref{gaugemassdiag}, and we also specify a constant
pseudo-abelian magnetic background field
\be
F^i_{\mu\nu}=\hat{m}^i F_{\mu\nu}\quad , \quad \hat{m}_i\hat{m}^i=1\quad , \quad F_{\mu\nu}=B\epsilon^\perp_{\mu\nu}\nonumber
\ee
where $\hat{m}$ is a unit vector pointing into a direction in the
Cartan of the algebra. The constant antisymmetric tensor
$\epsilon$ characterizes the space directions which are affected by
the constant magnetic field upon the Lorentz force. Recalling
that the adjoint generators are $(\tau^l)_{ij}=i f^{ilj}$, we choose a
basis in adjoint color space, such that $if^{ilj}\hat{m}^l$ is
diagonal with eigenvalues $\nu_i$. Then, the covariant
derivative
\be\label{diagadjD}
D_\mu^{ij}=(\partial_\mu-i\nu_i)\delta^{ij} \quad \text{(no sum over i)}
\ee
is also diagonal, and so are $D^2$ and ${\cal D}_\mathrm T$.
Hence ${\cal D}_\mathrm T$ and $\bar{m}_W^2$  commute, as well as $D^2$ and $\bar{m}_W^2$.
Equation~\eqref{totgaugetogauge} can thus be brought into propertime form:
\begin{align*}
{\rm STr}\left[\frac{\partial_t{R}_k}{\Gamma_k^{(2)}+R_k}\right]_{W}=
-\int_0^\infty \!\!ds\,\tilde{h}(s,0){\rm Tr}\left[e^{-s\frac{-D^2}{k^2}}\right]\\
+\int_0^\infty\!\!ds\,{\rm Tr}\left[\tilde{h}\!\left(s,m_W^2\right)\left(e^{-s\frac{{\cal D}_\mathrm T}{k^2}}-e^{-s\frac{- D^2}{k^2}}\right)\right]
\end{align*}
where $\tilde{h}$ is the Laplace transform of the function 
\be
h\!\left(y,m_W^2\right)=\frac{y\partial_t{r}_k(y)}{y(1+r_k(y))+m_W^2}\nonumber
\ee
with respect to $y$, that is
\be
h\!\left(y,m_W^2\right)=\int_0^\infty\!\!ds\,\tilde{h}\!\left(s,m_W^2\right)e^{-s y},\nonumber
\ee
where as before $m_W^2=\bar{m}^2_W/k^2$.
The heat kernel traces are known, see~\cite{Gies:2002af}
\begin{align}
\label{heatkernelforgauge}
{\rm Tr}\!\left[\tilde{h}\!\left(s,m_W^2\right)e^{-s\frac{{\cal D}_\mathrm T}{k^2}}\right]\!\!=
&\frac{\Omega k^4}{4\pi^2s^2}\!\sum_{i=1}^{\dG}\!\tilde{h}\!\left(s,m_{W,i}^2\right)  \nonumber\\
\times &\Big\{\frac{\frac{s b_i}{k^2}}{\sinh\!\left(\frac{s b_i}{k^2}\right)}+ \frac{s b_i}{k^2}\sinh\!\left(\frac{s b_i}{k^2}\right)\Big\} \nonumber\\
{\rm Tr}\! \left[\tilde{h}\! \left( s,m_W^2 \right) e^{ -s\frac{-D^2}{k^2} } \right]\!\!=
&\frac{\Omega k^4}{16\pi^2s^2}\!\sum_{i=1}^{\dG}\!\tilde{h}\!\left(s,m_{W,i}^2\right)
\frac{\frac{s b_i}{k^2}}{\sinh\!\left(\frac{s b_i}{k^2}\right)}\nonumber\\
{\rm Tr}\!\left[\tilde{h}(s,0)e^{-s\frac{-D^2}{k^2}}\right]\!\!=
&\frac{\Omega k^4}{16\pi^2s^2}\!\sum_{i=1}^{\dG}\!\tilde{h}(s,0)
\frac{\frac{s b_i}{k^2}}{\sinh\!\left(\frac{s b_i}{k^2}\right)}
\end{align}
where $b_i=\bar{g} |\nu_i| B$ and $\Omega$ is the spacetime
volume. The first trace above is over spacetime and Lorentz and color
indices, the other two only over spacetime and color indices.  For the
running gauge coupling we just need the terms of order $b_i^2$, since
the relevant term on the l.h.s. of the flow equation is
$\partial_t{\Gamma}_k\ni\Omega B^2\partial_t Z_W/2$. Using that the
running of the renormalized coupling $g^2$ is given in terms of the
anomalous dimension, $\partial_t g^2=\beta_{g^2}=\eta_W g^2$, we find
\be\label{etaWgauge}
\eta_W\Big|_W\!=\frac{-g^2}{32\pi^2}\!\sum_{i=1}^{\dG}\!\left[21h\!\left(0,m_{W,i}^2\right)+h(0,0)\right]\!\frac{|\nu_i|^2}{3}.
\ee
In the background-field method, the $y\rightarrow 0$ limit of the regulator is constrained
\cite{Pawlowski:2001df,Gies:2002af}; the only regulators permitted must
satisfy $h(y\rightarrow 0,0)=2$.  In the massless limit we thus obtain
\be
\eta_W\Big|_W=-\frac{1}{16\pi^2}\frac{22}{3}g^2\sum_{i=1}^{\dG}|\nu_i|^2=-\frac{1}{16\pi^2}\frac{22}{3}g^2
\NL,\nonumber 
\ee
which agrees with standard perturbation theory.  Let us work out the
massive case using the linear regulator~\eqref{eq:cutoff}. In this
case, $h(y,x)=2(1+x)^{-1}\theta(1-y)$, such that the gauge contribution to the
gauge $\beta_{g^2}$ function reeds
\be
\beta_{g^2}\Big|_W=\eta_W\Big|_W g^2=-\frac{g^4}{16\pi^2}\left[\frac{21}{3}\sum_{i=1}^{\dG}\frac{|\nu_i|^2}{1+m_{W,i}^2}+\frac{\NL}{3}\right]\,.\nonumber
\ee
The first term now depends on the choice of $\hat{n}^a$ the direction of the vev in fundamental color space. 
This is expected, as for higher gauge groups different breaking patterns and gauge masses
can arise. This term also depends in general on $|\nu_i|^2$, i.e. on $\hat{m}^i$. 
This is also plausible, as the directions of the vev implicitly also allows for the definition of different couplings:
depending on the relative direction of the gauge fluctuation w.r.t. the vev, the fluctuations can couple differently to matter.

For SU$(2)$, these issues simplify, as
\begin{equation}\label{gaugemassSU2}
\bar{m}_W^{2\ \ ij}\!\!=\!\frac{\bar{g}^2 \bar{v}^2}{4}\hat{n}^{\dagger a}\sigma^i_{ab}\sigma^j_{bc}\hat{n}^c
\!=\!\frac{\bar{g}^2 \bar{v}^2}{4}(\delta^{ij}\!+i\epsilon^{ijl}(\hat{n}^\dagger\sigma^l\hat{n})),
\end{equation}
such that ${\rm tr}\,\bar{m}_W^{2\ \ ij}\!=\!3\bar{g}^2 \bar{v}^2/4$.  Let us
denote $c^l=(\hat{n}^\dagger\sigma^l\hat{n})$. This is a vector in
adjoint space which is an eigenvector of the mass matrix, with
eigenvalue $\bar{g}^2 \bar{v}^2/4$.  One can choose a diagonalizing orthonormal
basis $\{e_1, e_2=c/|c|, e_3\}$ in adjoint space such that the
mass matrix takes the form
\be
\bar{m}_W^{2}=\frac{\bar{g}^2 \bar{v}^2}{4}\begin{pmatrix} 2&0&0\\0&1&0\\0&0&0\\\end{pmatrix}\,.\nonumber
\ee
Now recall that the $\nu_i$ denotes the eigenvalues of
$(-if^{ijl}\hat{m}^l)$, that for SU$(2)$ simply is
$(-i\epsilon^{ijl}\hat{m}^l)$.  Therefore in SU$(2)$ the eigenvalues
are $(1,-1,0)$ for any choice of $\hat{m}$. However, depending on the
direction of $\hat{m}$ w.r.t. the basis defined above, the $\nu_i$
could be $\{\nu_1=1, \nu_2=-1, \nu_3=0\}$ or possibly permutations
thereof. The two extreme cases for SU$(2)$ are maximal or minimal
decoupling.  Maximal decoupling happens if $|\nu_1|=|\nu_2|=1$ and
$\nu_3=0$, and in this case
\be\label{betamaximal}
\beta_{g^2}\Big|_W=-\frac{g^4}{16\pi^2}\left[\frac{21}{3}\left(\frac{1}{1+\frac{g^2 v^2}{2k^2}}+\frac{1}{1+\frac{g^2 v^2}{4k^2}}\right)+\frac{2}{3}\right]
\ee
while minimal decoupling happens if $\nu_1=0$ and $|\nu_2|=|\nu_3|=1$, and correspondingly
\be\label{betaminimal}
\beta_{g^2}\Big|_W=-\frac{g^4}{16\pi^2}\left[\frac{21}{3}\left(1+\frac{1}{1+\frac{g^2 v^2}{4k^2}}\right)+\frac{2}{3}\right]\,.
\ee
For SU$(2)$ the ambiguity of the $\beta$-function arises solely from
the ambiguity of defining a coupling in the presence of a vev.  In
fact, there are more quadratic invariants than the only $F^2$, such as
for example $\hat{n}^{\dagger
  a}F^i_{\mu\nu}T^i_{ab}T^j_{bc}F^j_{\mu\nu}\hat{n}^c$.  For higher
groups, even the mass matrix depends on the choice of $\hat{n}^a$.
\newline

\subsection{Contribution from scalar modes}

The contribution from scalar fluctuations to the gauge $\beta$
function arises from the scalar kinetic term.  The calculation is very
similar to that of the longitudinal gauge modes with two differences:
the field is complex and lives in the fundamental representation.
Moreover the dimensionless scalar mass matrix in the broken regime
reads $m_\phi^{2\,ab}=(\lambda_2 v^2/2k^2)\hat{n}^a\hat{n}^{\dagger
  b}$. Here, we do not attempt to solve the problem in full generality as
for the gauge modes, but confine ourselves to a simple choice of
backgrounds.  Most importantly, we choose the direction of the
pseudo abelian background to satisfy
\be\label{pseudoabelianback2}
W^i_\mu\!=\!\hat{m}^i W_\mu\ ,\quad\hat{m}_i\hat{m}^i\!=\!1\ ,\quad[(\hat{m}_iT^i),\hat{n}\otimes\hat{n}^\dagger]\!=\!0\,.
\ee
It is important to note that this does not constrain the choice of the
vev-direction $\hat{n}^a$.  This is because we can always choose a
basis in fundamental color space such that the projector
$P_{\hat{n}}=\hat{n}\otimes\hat{n}^\dagger$ is diagonal.  Then the
commutation relation~\eqref{pseudoabelianback2} can be satisfied by
choosing $(\hat{m}_iT^i)^{ab}$ to be in the Cartan, i.e. by choosing
it to be diagonal in that basis.

Let's consider SU$(2)$ as an example. Let $\hat{n}=(0,1)$.  Then we
choose $\hat{m}=(0,0,1)$ such that 
\be
(\hat{m}_iT^i)^{ab}=\frac{1}{2}\sigma^3=\frac{1}{2}
\begin{pmatrix}1&0\\0&-1\\\end{pmatrix},\label{eq:sccond}  
\ee 
obviously satisfying \Eqref{pseudoabelianback2}. Before we continue
with the scalar fluctuations, let us work out the consequences of this
choice for the gauge modes of the preceding section.  The vector $c$
for this choice becomes $c=(0,0,-1)$ and the mass matrix for the gauge
modes, given by~\eqref{gaugemassSU2}, is
\be\label{explicitgaugemassSU2}
\bar{m}_W^{2\ \ ij}=\frac{\bar{g}^2
  \bar{v}^2}{4}(\delta^{ij}-i\epsilon^{ij3})
=\frac{\bar{g}^2 \bar{v}^2}{4}\begin{pmatrix} 1&-i&0\\i&1&0\\0&0&1\\\end{pmatrix}.
\ee
The definition of $\nu_i$, right above~\eqref{diagadjD}, combined with
the choice $\hat{m}=(0,0,1)$ requires us to compute the eigenvalues of
\be
-i\epsilon^{ij3}=-i\begin{pmatrix} 0&1&0\\-1&0&0\\0&0&0\\\end{pmatrix}\nonumber\,.
\ee
The simultaneous eigenvectors of this matrix and of $\bar{m}_W^2$ are
given by
\be
v_1=\begin{pmatrix} 0\\0\\1\\\end{pmatrix} \quad,\quad v_2=\begin{pmatrix} 1\\i\\0\\\end{pmatrix} \quad,\quad v_3=\begin{pmatrix} 1\\-i\\0\\\end{pmatrix}\nonumber
\ee
with the corresponding set of eigenvalues:
$\{\bar{m}_{W,1}^2=\frac{\bar{g}^2 \bar{v}^2}{4},\nu_1=0\}$,
$\{\bar{m}_{W,2}^2=\frac{\bar{g}^2 \bar{v}^2}{2},\nu_2=1\}$,
$\{\bar{m}_{W,3}^2=0,\nu_3=-1\}$.  This choice of $\hat{m}$
corresponds to the minimal decoupling case of \Eqref{betaminimal}.
These considerations tell us that the maximal decoupling solution of
\Eqref{betamaximal} might not be permitted, as it would not correspond
to a legitimate choice of $\hat{m}$ with $\hat{m}_iT^i$ in the Cartan
(which we had also assumed in the gluonic case in
eq.~\eqref{diagadjD}). The choice \eqref{pseudoabelianback2} for
defining $\hat{m}$ therefore is related to defining the coupling with
respect to the unbroken part of the gauge group.

Let us now return to the scalar fluctuations;
Eq.~\eqref{pseudoabelianback2} ensures that the covariant derivative
in the fundamental representation satisfies
\be
[D_\mu,\hat{n}\otimes\hat{n}^\dagger]=0\nonumber
\ee
for our choice of the background field. Then also $[-D^2,\hat{n}\otimes\hat{n}^\dagger]=0$ and thus $[-D^2,m_\phi^2]=0$ follow, such that $-D^2$ and $m_\phi^2$ can be simultaneously
diagonalized. Therefore
\be
{\rm Tr}\left[\frac{\partial_t{R}_k}{\Gamma^{(2)}+R_k}\right]_{\phi}=
{\rm Tr}\left[\frac{\frac{-D^2}{k^2}\partial_t{r}_k\!\left(\frac{-D^2}{k^2}\right)}{\frac{-D^2}{k^2}\left(1+r_k\!\left(\frac{-D^2}{k^2}\right)\right)+m_\phi^2}\right]\,.\nonumber
\ee
Because of the above considerations, we can rewrite the previous
expression in the propertime form
\begin{align*}
{\rm Tr}\left[\frac{\partial_t{R}_k}{\Gamma^{(2)}+R_k}\right]_{\phi}&=
\int_0^\infty\!\! ds\, {\rm Tr}\left[\tilde{h}(s,m_\phi^2)e^{-s\frac{-D^2}{k^2}}\right]\\&=\frac{\Omega}{16\pi^2}
\int_0^\infty\!\! ds\, \sum_{a=1}^{\NL}\tilde{h}(s,m_{\phi,a}^2)\left(-\frac{1}{6}b_a^2\right),
\end{align*}
where we have retained only the term of order $b_a^2$ (compare with the second
equation of~\eqref{heatkernelforgauge}). We have denoted the
eigenvalues of the mass matrix by $m_{\phi,a}^2$ (there is only one
nonvanishing eigenvalue for the radial mode). Furthermore,
$b_a=\bar{g}|\nu_a|B$, where $\nu_a$ now are the eigenvalues of
$\left(\hat{m}_iT^i\right)^{ab}$ related to the fundamental
representation.  Using the standard normalization for
the generators of the fundamental representation, we have
\be
\sum_{a=1}^{\NL}|\nu_a|^2={\rm tr}\left[\left(\hat{m}_iT^i\right)^2\right]
=\hat{m}_i\hat{m}_j\frac{1}{2}\delta^{ij}=\frac{1}{2}\,.\nonumber
\ee
Another difference from the gauge case is that the scalar field is
complex and thus there is no factor $1/2$ in front of the trace on the
r.h.s. of the flow equation.  Hence, analogous to~\eqref{etaWgauge},
the contribution of the scalar to the flow of $Z_W$ reads
\be\label{etaFscalar}
\eta_W\Big|_{\phi}=\frac{g^2}{16\pi^2}\sum_{a=1}^{\NL}h(0,m_{\phi,a}^2)\frac{|\nu_a|^2}{3}\,.
\ee
In the massless case, since $h(0,0)=2$, $
\eta_W|_{\phi}=\frac{g^2}{16\pi^2}\frac{1}{3}\nonumber $ in agreement
with perturbation theory.  In the general massive case and using the
linear regulator, we get
\be
\eta_W\Big|_{\phi}=\frac{g^2}{16\pi^2}\frac{2}{3}\sum_{a=1}^{\NL}\frac{1}{1+m_{\phi,a}^2}|\nu_a|^2\nonumber\,.
\ee
Generically, only one particular component of $m_{\phi,a}^2$ is
nonvanishing and equal to $2\lambda_2\kappa$.  For SU$(2)$ the $\nu_a$
are unique and equal to $\{\frac{1}{2},-\frac{1}{2}\}$.  Therefore in
this case
\be
\eta_W\Big|_{\phi}=\frac{g^2}{16\pi^2}\frac{1}{3}\left[\frac{1}{2}+\frac{1}{2}\frac{1}{1+2\lambda_2\kappa}\right]\nonumber\,.
\ee
%

\subsection{Contribution from fermion modes}

The relevant part of the effective Lagrangian is
\be
{\cal L}_k \ni i(\bar{\psi}_\mathrm L^a\slashed{D}^{ab}\psi_\mathrm L^b+\bar{\psi}_\mathrm R\slashed{\partial}\psi_\mathrm R)
+\bar{h}(\bar{\psi}_\mathrm R\phi^{a\dagger}\psi_\mathrm L^a-\bar{\psi}_\mathrm L^a\phi^{a}\psi_\mathrm R\nonumber),
\ee
where again we have set any wave function renormalization to one.
For~\Eqref{fluctandvev}, we can choose a gauge background field such
that $D_\mu^{ab}$ and $P_{\hat{n}}=\hat{n}\otimes\hat{n}^\dagger$ as
well as $P_{(1-\hat{n})}=1-P_{\hat{n}}$ commute and the above
parts of ${\cal L}_k$ can be written as
\begin{align}\label{Lforfermionstogauge}
{\cal L}_k& \ni i(\bar{\psi}_\mathrm L^a\slashed{D}^{ab}P_{(1-\hat{n})}^{bc}\psi_\mathrm L^c)\\
&+i(\bar{\psi}_{\mathrm L}^{\hat{n}}\slashed{D}\psi_{\mathrm L}^{\hat{n}}+\bar{\psi}_\mathrm R\slashed{\partial}\psi_\mathrm R)
+\frac{\bar{h} \bar{v}}{\sqrt{2}} ( \bar{\psi}_\mathrm R\psi_{\mathrm L}^{\hat{n}}-\bar{\psi}_{\mathrm L}^{\hat{n}}\psi_\mathrm R)\nonumber
\end{align}
Here, the $\slashed{D}$ in the second term is projected along ${\hat
  n}$.  The first line corresponds to the massless bottom-type
fermions. Their contribution is the standard perturbative contribution
weighted by eigenvalues $\nu_a$ in the orthogonal complement.  Let
$\hat{n}$ point into the $A$-direction: $\hat{n}^a=\delta^{aA}$. Then
the contribution of the massless fermions to the running coupling is
\be
\partial_tg^2\Big|_{\psi_{(1-\hat{n})}}=\frac{g^4}{16\pi^2}\frac{2d_\gamma \Ng}{3}\sum_{a=1,a\neq A}^{\NL} |\nu_a|^2\,.\nonumber
\ee
If the sum ran over all $a$'s we would get $\sum_{a=1}^{\NL}
|\nu_a|^2=1/2$ leading to the correct perturbative result.  Combining
$\psi_{\mathrm L}^{\hat{n}}$ and $\psi_\mathrm R$ into a Dirac spinor
$\Psi=\begin{pmatrix} \psi_{\mathrm L}^{\hat{n}}\\\psi_\mathrm
R \end{pmatrix}$, the second line of~\eqref{Lforfermionstogauge} can
be written
\be
{\cal L}_k \ni i\bar{\Psi}\slashed{D}_{A\mathrm L}\Psi+\bar{m}_t \bar{\Psi}\gamma_5\Psi\nonumber
\ee
where $\slashed{D}_{A\mathrm L}=\gamma_\mu(\partial^\mu-\bar{g}\nu_A
W^\mu P_\mathrm L)$, with the usual definition of the left projector
$P_\mathrm L=\frac{1}{2}(1-\gamma_5)$. We have also introduced the
``top-mass'' $\bar{m}_t$ as defined in eq.~\eqref{eq:topmass}.
Since the regularized fluctuation operator for $\Psi$ satisfies
$(\Gamma^{(2)}_k+R_k)=\slashed{D}_{A\mathrm
  L}(1+r_k)+\bar{m}_t\gamma_5$ and since ${\rm
  tr}[\gamma_5\slashed{D}_{A\mathrm L}]=0$, we get
\begin{align}\label{temporarytoptogauge}
{\rm Tr}\left[\frac{\partial_t{R}_k}{\Gamma^{(2)}+R_k}\right]_{\Psi}&={\rm Tr}\left[
\frac{
\slashed{D}_{A\mathrm L}^2\left(1+r_k\!\left(\frac{\slashed{D}_{A\mathrm L}^2}{k^2}\right)\right)\partial_t{r}\!\left(\frac{\slashed{D}_{A\mathrm L}^2}{k^2}\right)
}{
\slashed{D}_{A\mathrm L}^2\left(1+r_k\!\left(\frac{\slashed{D}_{A\mathrm L}^2}{k^2}\right)\right)^2+\bar{m}_t^2
}\right]\nonumber\\
&=\int_0^\infty\!\!ds\,\tilde{h}(s,\bar{m}_t^2){\rm Tr}\left[e^{-s\frac{\slashed{D}_{A\mathrm L}^2}{k^2}}\right]
\end{align}
Here we need to know the spectrum of 
\begin{widetext}
\begin{eqnarray}\label{DALdiffersfromDL}
\slashed{D}_{A\mathrm L}^2&=&\gamma_\mu(\partial^\mu-\bar{g}\nu_A W^\mu P_\mathrm L)\gamma_\nu(\partial^\nu-\bar{g}\nu_A W^\nu P_\mathrm L)=
\gamma_\mu\gamma_\nu(\partial^\mu-\bar{g}\nu_A W^\mu P_\mathrm R)(\partial^\nu-\bar{g}\nu_A W^\nu P_\mathrm L)\nonumber\\
&=&\gamma_\mu\gamma_\nu(D_\mathrm R^\mu+\partial_\mathrm L^\mu)(D_\mathrm L^\nu+\partial_\mathrm R^\nu)=
\slashed{D}_{\mathrm L}^2+\gamma_\mu\gamma_\nu(\partial_\mathrm L^\mu D_\mathrm L^\mu+D_\mathrm R^\mu\partial_\mathrm R^\nu)
\end{eqnarray}
\end{widetext}
where we have denoted $\partial_{\mathrm{L/R}}^\mu=\partial^\mu
P_{\mathrm{L/R}}$ and took advantage of: $\slashed{D}_{\mathrm
  L}^2=\gamma_\mu\gamma_\nu D_\mathrm R^\mu D_\mathrm L^\nu$ and
$\partial_\mathrm L^\mu\partial_\mathrm R^\nu=0$.  The determination
of this spectrum probably is an analytically soluble problem for a
constant magnetic field background, as the differential operator is of
harmonic oscillator type, however, with an involved Dirac structure. 

As we are mainly interested in the decoupling of massive modes in the
flow of the gauge coupling, let us simply take a shortcut at this
point. We already know that the contribution
of~\Eqref{temporarytoptogauge} to the $\beta$-function in the massless
limit must be of the form
\be\label{masslesstop}
\partial_tg^2\Big|_{\psi_{\hat n}}=\frac{g^4}{16\pi^2}\frac{2d_\gamma \Ng}{3} |\nu_A|^2\,.\nonumber
\ee
This fixes the ${\cal O}(s^0)$-term in ${\rm
  Tr}\left[e^{-s\frac{\slashed{D}_{A\mathrm L}^2}{k^2}}\right]$ to be
the same as the ${\cal O}(s^0)$-term in ${\rm
  Tr}\left[e^{-s\frac{\slashed{D}_{\mathrm L}^2}{k^2}}\right]$.  These
heat-kernel traces could still differ to higher orders in $s$, due to
the two extra terms in~\eqref{DALdiffersfromDL}.  These higher-order
terms could (unlike as for $\slashed{D}_{\mathrm L}^2$) in principle
contain terms of order $B^2$ and thus contribute to the beta function
via functions of the form
\be
f_p(m_t^2)=\int_0^\infty\!\!ds\,\tilde{h}(s,m_t^2)s^p=\left[\left(-\frac{\partial\ }{\partial y}\right)^p h(y,m_t^2)\right]_{y=0}\nonumber
\ee
where $m_t^2$ is the dimensionless top mass squared:
$m_t^2=h^2\kappa$.  Because of the constraints from the
massless limit discussed before, we must have $f_p(0)=0$. Furthermore
$f_p$ also has to exhibit a generic threshold behavior, that is:
$f_p(m_t^2\rightarrow\infty)\rightarrow 0$.  As the precise
dependence of $h(0,m_t^2)$ is anyway regulator-dependent, we
simply ignore potentially nonvanishing contributions of
$f_p(m_t^2)$ for all practical discussions in the main text.
Therefore, without any further explicit calculation, we approximate
the threshold behavior of the massive fermion mode by the same form as
for the other modes
\be
\partial_tg^2|_{\psi}=\frac{g^4}{16\pi^2}\frac{2d_\gamma \Ng}{3}\sum_{a=1}^{\NL}
\frac{1}{1+h^2\kappa\delta^{aA}}|\nu_a|^2\,.\nonumber 
\ee
For SU$(2)$ this implies
\be
\partial_tg^2|_{\psi}=\frac{g^4}{16\pi^2}\frac{d_\gamma \Ng}{3}\left(\frac{1}{2}+\frac{1}{2}\frac{1}{1+h^2\kappa}\right)\,.\nonumber
\ee
\newline

To summarize, we can write the gauge one-loop $\beta$-function approximately as given in the main text
in eqs.~(\ref{floweq:gauge},\ref{floweq:gaugeLSU2}).

\section{Results for the anomaly-free two-generation model $\Ng=2$}
\label{sec:Ng2}

\begin{figure}[t]
\begin{center}
\includegraphics[width=0.23\textwidth]{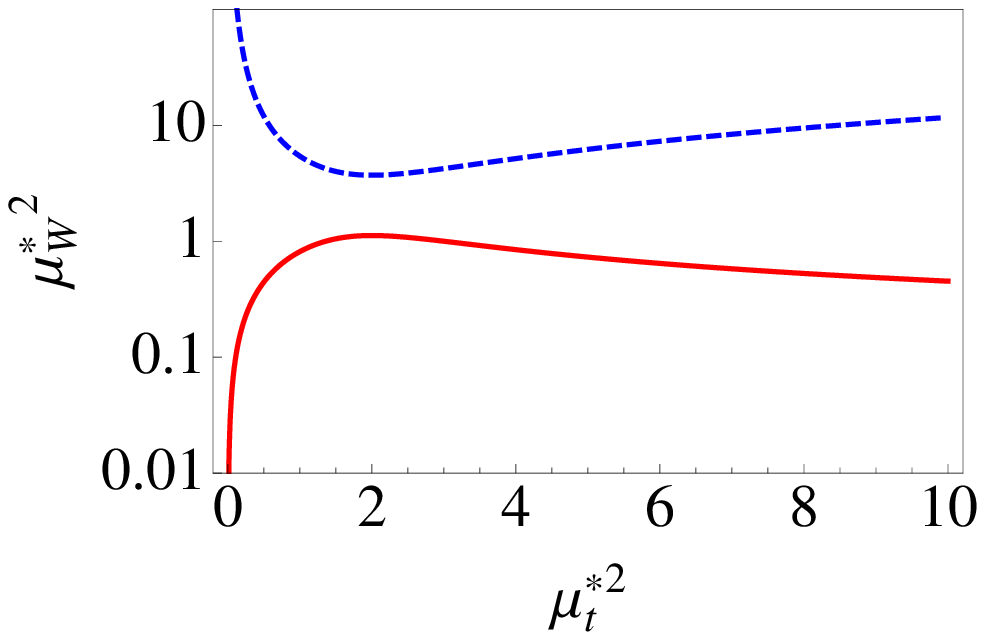}
\includegraphics[width=0.23\textwidth]{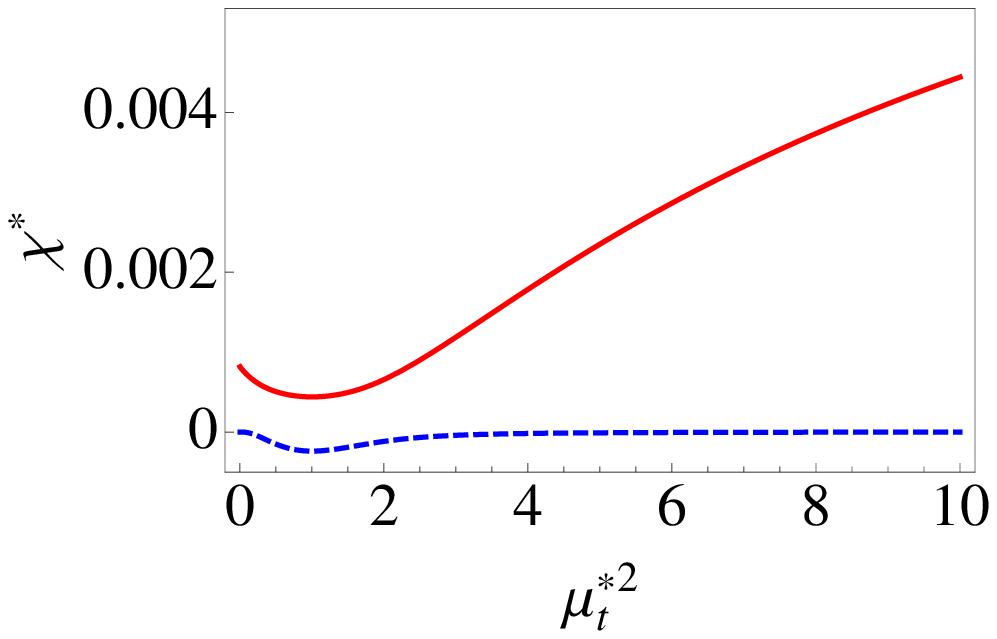}
\caption{Fixed point values for $\mu_W^2$ (left panel) and $\chi$ (right
  panel) as a function of the fixed point value of $\mu_{\mathrm t}^2$ for
  $\NL=2$ and $N_g=2$. This shows the similarity to the one-generation case
  depicted in Fig.~\ref{fig:fixedpointsNL2}.}
\label{twogen1}
\end{center}
\end{figure}

\begin{figure}[t]
\begin{center}
\includegraphics[width=0.23\textwidth]{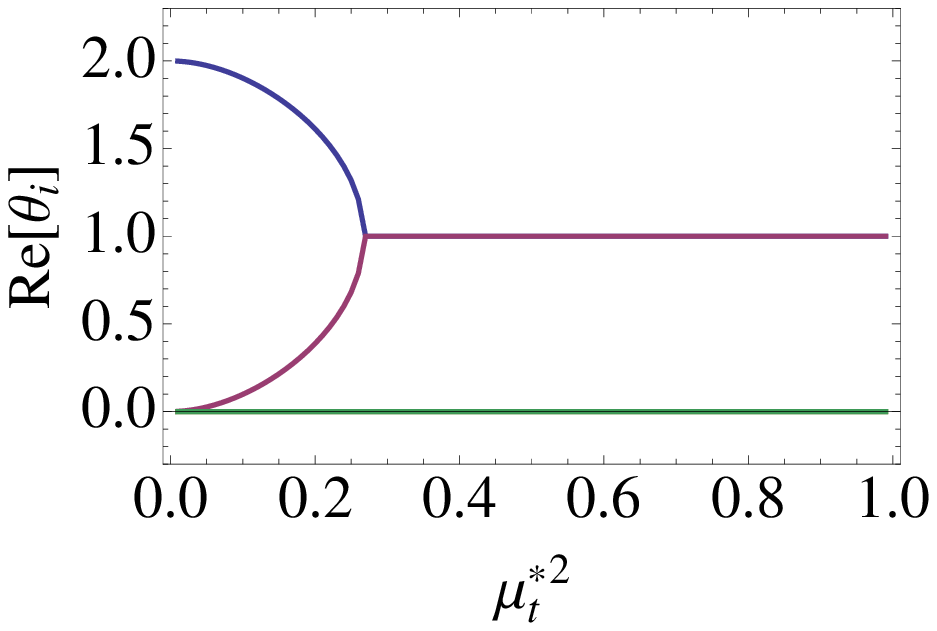}
\includegraphics[width=0.23\textwidth]{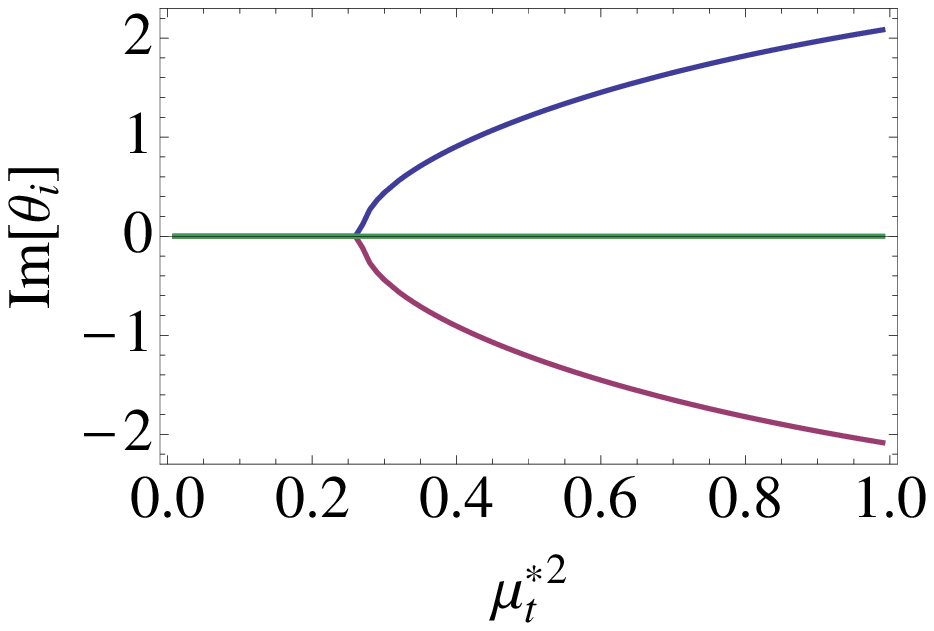}
\caption{Critical exponents for the line of fixed points computed in the mass
  parametrization as a function of the fixed point top mass parameter
  $\mu_{\text{t}}^{\ast 2}$ for $\NL=2$ and $N_g=2$; left panel: real parts,
  right panel: imaginary parts. Again the similarity to the one-generation
  case, c.f., Fig.~\ref{fig:thetaNL2} is obvious.}
\label{twogen2}
\end{center}
\end{figure}

In this appendix, we verify explicitly that the properties of the anomaly-free
SU($\NL=2$) model with two left-handed generations $\Ng = 2$ are essentially
identical to the results for the one-generation model discussed in
the main text. This can be seen manifestly by comparing Figs.~\ref{twogen1},
\ref{twogen2}, where $\Ng = 2$, to Figs.~\ref{fig:fixedpointsNL2},
\ref{fig:thetaNL2}, where $\Ng = 1$. These explicit solutions verify that the
transition from one generation to two generations induces only small
quantitative differences in the fixed-point values as well as in the values
for the critical exponents.

We emphasize again that the $\Ng=1$ model discussed in the main text for
phenomenological reasons has a Witten anomaly and thus should be considered as
embedded into a larger anomaly free model, such as the standard
model. By contrast, with the results of this appendix, we conclude that the
$\Ng=2$ model as it stands can be a consistent UV complete quantum field
theory for all trajectories emanating from the line of non-Gau\ss ian fixed
points.

\end{document}